\documentclass{appolb}
\usepackage{epsfig}
\begin{document}
\title{Diffraction at the Tevatron and the LHC
\thanks{Presented at the Summer School on QCD, low x physics, and Diffraction 
in Copanello, Calabria, Italy, 1-14 July 2007.}}
\author{C. Royon
\address{IRFU/Service de physique des particules \\ CEA/Saclay, 91191 
Gif-sur-Yvette cedex, France}}
\maketitle
\begin{abstract}
In these lectures, we present and discuss the most recent results on
inclusive diffraction at the Tevatron collider and give the
prospects at the LHC. We also describe the search for exclusive events
at the Tevatron. Of special interest is the exclusive
production of Higgs boson and heavy objects ($W$, top, stop pairs) at the LHC
which will
require precise measurements and analyses of inclusive and exclusive
diffraction to constrain further the gluon density in
the pomeron. At the end of these lectures, we describe the projects to install
forward detectors at the LHC to fulfil these measurements.
\end{abstract}

In these lectures, we describe the most recent results on inclusive
diffraction at the Tevatron, and especially the search for exclusive events. 
We finish the lecture by discussing the prospects of diffractive physics at the LHC,
and in particular the exclusive diffractive Higgs production. We also describe the
diffractive experiments 
accepted or in project at the LHC: TOTEM, ALFA in ATLAS, and the AFP/FP420 projects.

\section{Experimental methods to select diffractive events at the Tevatron
and the LHC}

In this section, we discuss the different experimental ways to define
diffraction at the Tevatron and the LHC. 

The Tevatron is a $p \bar{p}$ collider located close to Chicago at Fermilab,
USA. It is presently the collider with the highest center-of-mass energy of
about 2 TeV. Two main experiments are located around the ring, D\O\ and CDF. 
Both collaborations have accumulated a luminosity larger than 3. fb$^{-1}$ with an efficiency
larger than 90\%.

As a starting point, we describe the methods to select diffractive events
used by the H1 and ZEUS
experiments at HERA, DESY, Hamburg in Germany since it is easier. For more details about 
diffraction at HERA, see the
lectures from Bernd Loehr at this Summer school~\cite{loehr}. 

\subsection{The rapidity gap method}
\begin{figure}
\begin{center}
\vspace{10.cm}
\hspace{-5cm}
\epsfig{file=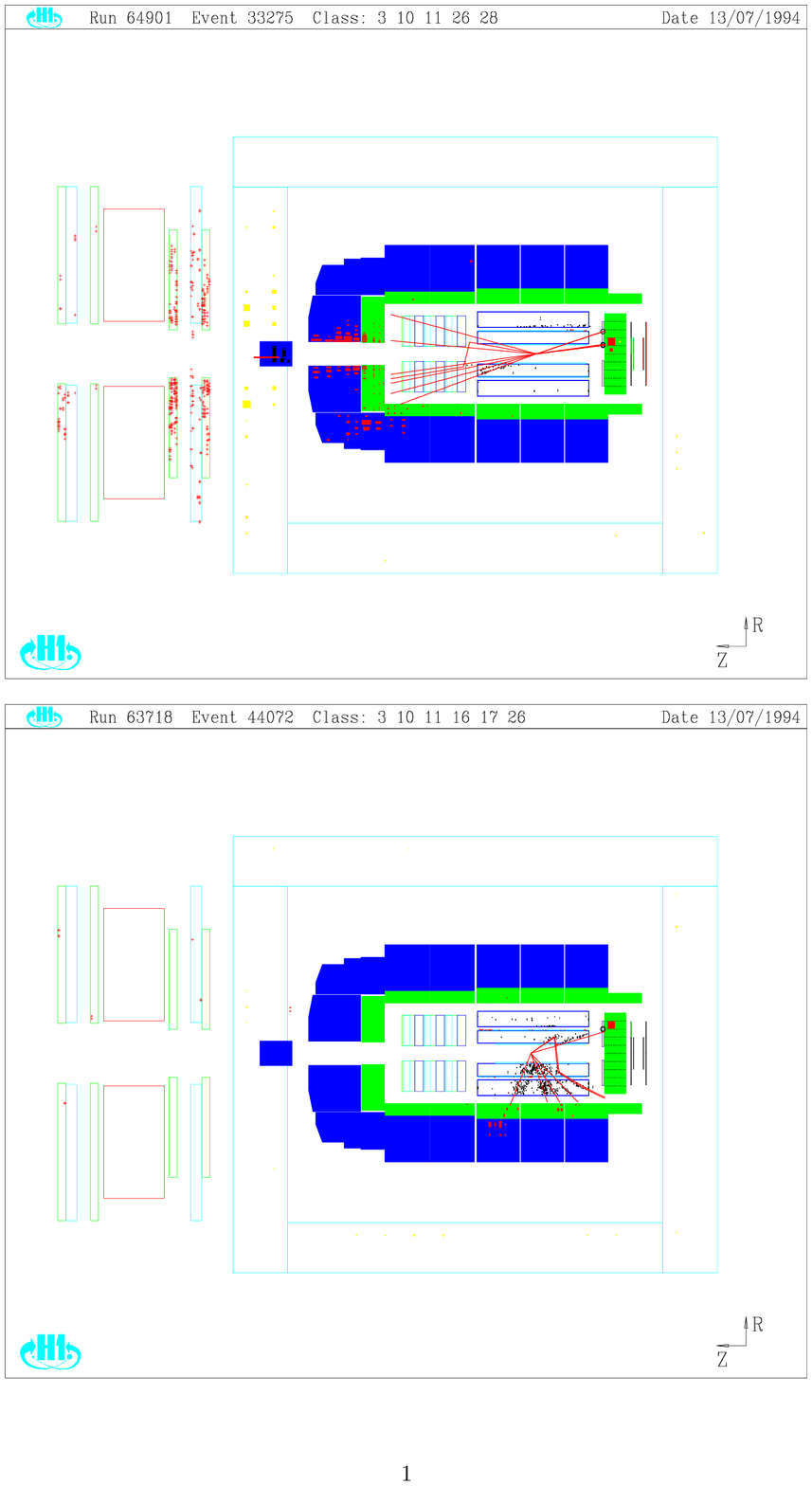,width=5.2cm}
\caption{``Usual" and diffractive events in the H1 experiment.}
\label{fig1}
\end{center}
\end{figure}

HERA was a collider where electrons of 27.6 GeV collided with protons of 920 GeV.
A typical event as shown in the upper plot of Fig.~\ref{fig1} is $ep \rightarrow eX$
where electron and jets are produced in the final state. We
notice that the electron is scattered in the H1 backward detector\footnote{At
HERA, the backward (resp. forward) directions are defined as the direction
of the outgoing electron (resp. proton).} (in green)
whereas some hadronic activity is present in the forward region of the detector
(in the LAr calorimeter and in the forward muon detectors). The proton is thus
completely destroyed and the interaction leads to jets and proton remnants directly observable
in the detector. The fact that much energy is observed in the forward region is
due to colour exchange between the scattered jet and the proton remnants.
In about 10\% of the events, the situation is completely
different. Such events appear like the one shown in the bottom plot of Fig.~\ref{fig1}.
The electron is still present in the backward detector, there is
still some hadronic activity (jets) in the LAr calorimeter, but no energy above
noise level is deposited in the forward part of the LAr calorimeter or in the
forward muon detectors. In other words, there is no color exchange between the
proton and the produced jets. As an example, this can be explained if the proton stays intact
after the interaction. 

This experimental observation leads to the first definition of diffraction:
request a
rapidity gap (in other words a domain in the forward detectors where  no
energy is deposited above noise level) in the forward region. For example, the
D0 and CDF 
collaborations at the Tevatron, as well as H1 and ZEUS at HERA,
use this method to select diffractive events as we will see it in the following.
Let us note that this
approach does not insure that the proton stays intact after the interaction, but 
the proton could be dissociated. The advantage of the rapidity gap
method is that it is quite easy to implement and it has a large acceptance in 
the diffractive kinematical
plane.

\subsection{Proton tagging}
The second experimental method to detect diffractive events is also natural: the
idea is to detect directly the intact proton in the final state. The proton
loses a small fraction of its energy and is thus scattered at very small angle
with respect to the beam direction. Some special detectors called roman pots can
be used to detect the protons close to the beam. The basic idea is simple: the roman pot
detectors are located far away from the interaction point and can move close to
the beam, when the beam is stable, to detect protons scattered at vary small
angles. The inconvenience is that the kinematical reach of those detectors is
much smaller than with the rapidity gap method, especially at HERA. On the other hand,
the advantage is that it
gives a clear signal of diffraction since it measures the diffracted proton
directly. This method is also used at the Tevatron and at HERA, and such detectors 
are or will be installed at the LHC.

\begin{figure}
\begin{center}
\epsfig{file=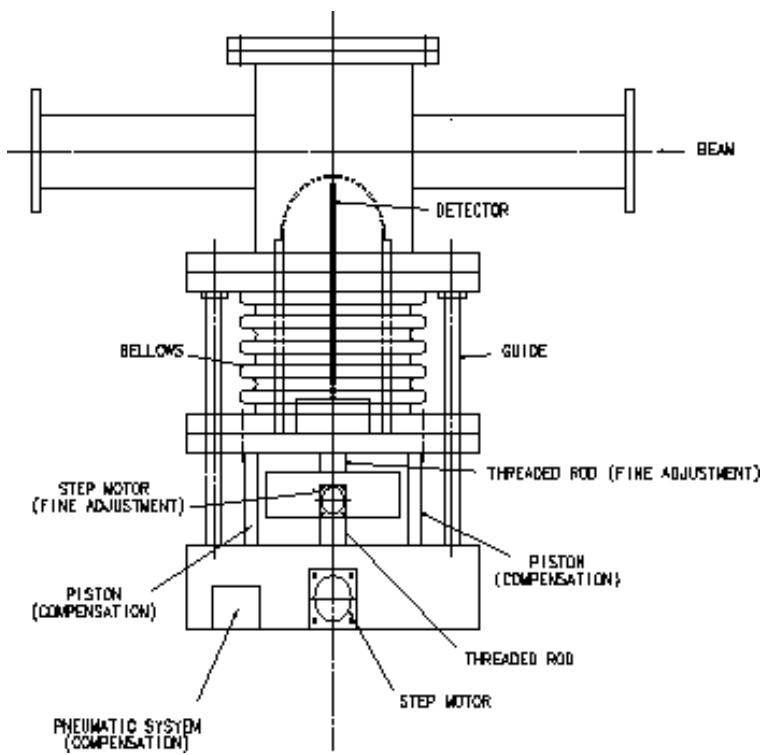,width=10cm}
\caption{Scheme of a roman pot detector.}
\label{fig2}
\end{center}
\end{figure}

A scheme of a roman pot detector as it is used by the H1 or ZEUS experiment is shown
in Fig.~\ref{fig2}. The beam is the horizontal line at the upper part of the
figure. The detector is located in the pot itself and can move closer to the
beam when the beam is stable enough (during the injection period, the detectors
are protected in the home position). Step motors allow to move the detectors
with high precision. A precise knowledge  of the detector position is
necessary to reconstruct the transverse momentum of the scattered proton and
thus the diffractive kinematical variables. The detectors are placed in a
secondary vaccuum with respect to the beam one. 

\subsection{Diffractive kinematical variables}

The difference between diffraction at HERA~\cite{loehr} and at the Tevatron is that
diffraction can occur not only on either $p$ or $\bar{p}$ side as at 
HERA, but also on both sides. The former case is called single diffraction
whereas the other one double pomeron exchange. In the same way
as the kinematical variables $x_P$ and $\beta$ are defined
at HERA, we define $\xi_{1,2}$(=$x_P$ at HERA) 
as the proton fractional momentum loss (or as the $p$ or
$\bar{p}$ momentum fraction carried by the pomeron), and $\beta_{1,2}$, the fraction of the
pomeron momentum carried by the interacting parton. The produced diffractive
mass is equal to $M^2= s \xi_1 $ for single diffractive events and to
$M^2= s \xi_1 \xi_2$ for double pomeron exchange where
$\sqrt s$ is the center-of-mass energy~\footnote{This formula is valid when the
mass of the produced object is larged compared to the proton mass for instance.
A more detailed formula is available in Ref.~\cite{chic} when this is not the 
case.}. The size of the rapidity gap
is of the order of $\Delta \eta \sim \log 1/ \xi_{1,2}$.

\section{Results on inclusive diffraction from the Tevatron}

\subsection{Diffractive events at the Tevatron}
The D\O\ and CDF collaborations obtained
their first diffractive results using the rapidity gap method which showed that
the percentage of single diffractive events was of the order of 1\%, and about
0.1\% for double pomeron exchanges. This study was made for different experimental
observables ($Z$, jets, $J/\Psi$...) and was found to be always of the same order of
magnitude~\cite{oldtev}.
Unfortunately, the reconstruction of the
kinematical variables is less precise than at HERA if one uses the rapidity gap
selection since it suffers from the worse resolution of reconstructing hadronic
final states, and this method is not practical to obtain quantitative results.

The other more precise method is to tag directly the $p$ and $\bar{p}$ in the
final state, as we mentionned already. 
The CDF collaboration installed roman pot detectors in the outgoing
$\bar{p}$ direction only at the end of Run I~\cite{cdfpots}, whereas the D\O\ collaboration
installed them both in the outgoing $p$ and $\bar{p}$ directions~\cite{d0pots}. 
The D\O\ (dipole detectors) and CDF roman
pots cover the acceptance of $t$ close to 0 and $0.02 < \xi < 0.05$ in the
outgoing $\bar{p}$ direction only. In addition, the D\O\ coverage extends for
$0.5 < |t| < 1.5$ GeV$^2$, and $0.001< \xi < 0.03$ in both $p$ and $\bar{p}$
directions (quadrupole detectors).
The CDF collaboration completed the detectors in the forward region by adding a
miniplug calorimeter on both $p$ and $\bar{p}$ sides allowing a coverage of
$3.5 < |\eta| < 5.1$ and some beam showing counters close to beam pipe ($5.5 <
|\eta| < 7.5$) allowing to reject non diffractive events.

\subsection{From HERA to Tevatron}
\begin{figure}
\begin{center}
\epsfig{file=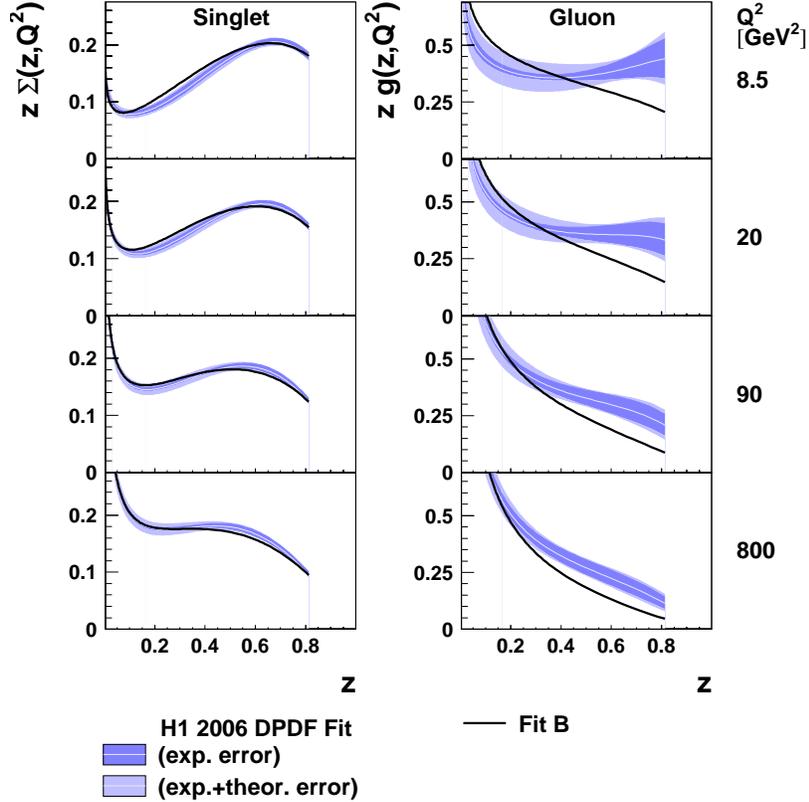,width=11cm}
\vspace{12cm}
\caption{Extraction of the parton densities in the pomeron using a DGLAP NLO fit
(H1 collaboration).}
\label{gluonh1}
\end{center}
\end{figure}

The starting point for the factorisation studies between HERA and the Tevatron is the
measurement of the diffractive structure function $F_2^D$ at HERA~\cite{f2d}. Many
different models are used to describe the diffractive structure functions (two-gluon
model~\cite{bekw}, dipole model~\cite{dipole}, or saturation
model~\cite{saturation}), which are described in another lecture~\cite{loehr}.
Here we will concentrate on the Dokshitzer Gribov Lipatov Altarelli Parisi  
(DGLAP)~\cite{dglap} fits to $F_2^D$ data. If we assume that the
pomeron is made of quarks and gluons~\cite{ingelman}, it is natural to check whether the DGLAP
evolution equations are able to describe the $Q^2$ evolution of these parton
densities. As necessary for DGLAP fits, a form for the input distributions is assumed
at a given $Q_0^2$ and is evolved using the DGLAP evolution equations to a
different $Q^2$, and fitted to the diffractive structure function data at
this $Q^2$ value~\cite{loehr,us}. The DGLAP QCD fit allows to get the parton distributions 
in the pomeron as a
direct output of the fit. The quark and gluon densities in the pomeron
for a $Q^2$ value of 8.5, 20, 90 and 800 GeV$^2$ are 
displayed in Fig.~\ref{gluonh1} as a function of $\beta$. The uncertainty
is displayed as a grey shaded
area around the central value. We first note that the gluon density is much
higher than the quark one, showing that the pomeron is gluon dominated. We also
note that the gluon density at high $\beta$ is poorly constrained which is shown
by the larger shaded area~\cite{us}. Another fit using another form of input
distribution is also displayed in the same figure as a black line~\cite{loehr}
which shows even more the bad determination of the gluon density in the pomeron
at high $\beta$.
Let us note that it is possible to constrain
further the gluon density in the pomeron by using in addition diffractive jet data in
the fit~\cite{loehr,f2d}. 
Even after using jet data, the uncertainty is still large at high $\beta$ ($\sim$50\%)
and we will see that this will have some consequences when we will discuss the search
for exclusive events at the Tevatron or the LHC.

Once the gluon and quark densities in the pomeron are known, it is easy to make
predictions for the Tevatron (or the LHC) if one assumes that the same mechanism is
the origin of diffraction in both cases. We assume the same structure of the pomeron
at HERA and the Tevatron and we compute as an example the jet production
in single diffraction or double pomeron exchange using the 
parton densities in the pomeron measured at HERA~\cite{us}. The interesting point is
to see if this simple argument works or not, or in other word if the factorisation
property between HERA and the Tevatron --- using the same parton distribution
functions --- holds or not~\cite{us}.

\subsection{Factorisation or factorisation breaking at the Tevatron?}

The CDF collaboration measured diffractive events at the Tevatron and their
characteristics. In general, diffractive events show as expected less QCD
radiation: for instance, dijet events are more back-to-back or the difference in
azimuthal angles between both jets is more peaked towards $\pi$. To make quantitative
predictions at the Tevatron and the LHC, it is useful to know if factorisation
holds as we mentioned in the previous section. In other words, is it possible to 
use the parton distributions in the pomeron 
obtained at HERA to make predictions at the Tevatron, and
also further constrain the parton distribution functions in the pomeron since
the reach in the diffractive kinematical plane at the Tevatron and HERA is
different? Theoretically, factorisation is not expected to hold between the
Tevatron and HERA~\cite{collins} due to additional $pp$ or $p \bar{p}$ interactions. 
For instance, some soft gluon
exchanges between protons can occur at a longer time scale than the hard
interaction and destroy the rapidity gap or the proton does not remain intact
after interaction. The factorisation break-up is confirmed by comparing the percentage of
diffractive events at HERA and the Tevatron (10\% at HERA and about 1\% of
single diffractive events at the Tevatron) showing already that factorisation
does not hold. This introduces the concept of gap survival probability, the
probability that there is no soft additional interaction or in other words that
the event remains diffractive. We will mention in the following how this concept
can be tested directly at the Tevatron for instance.

\begin{figure}
\begin{center}
\epsfig{file=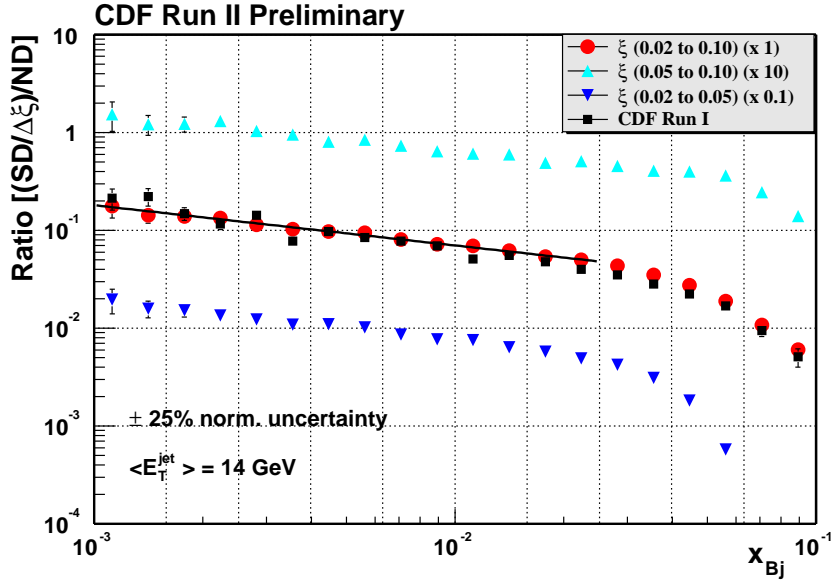,width=12cm}
\caption{Test of factorisation within CDF data alone.}
\label{fig3}
\end{center}
\end{figure}

The first experimental test of factorisation concerns CDF data
only. It is interesting to check whether factorisation holds within CDF data
alone, or in other words if the $\beta$ and $Q^2$ dependence can be factorised out
from the $\xi$ one. Fig.~\ref{fig3} shows the percentage of diffractive events
as a function of $x$ for different $\xi$ bins and shows the same $x$-dependence within
systematic and statistical uncertainties
in all $\xi$ bins supporting the fact that CDF data are consistent with 
factorisation~\cite{cdfdiff}. The CDF collaboration also studied the $x$ dependence for
different $Q^2$ bins which leads to the same conclusions. This also shows that
the Tevatron data do not require additional secondary reggeon trajectories as in
H1~\cite{loehr}. These results show that the soft interactions occuring at a
much longer time scale do not depend on the hard scattering.
It will be interesting to check if the same conclusions hold at the LHC.
 
\begin{figure}
\begin{center}
\epsfig{file=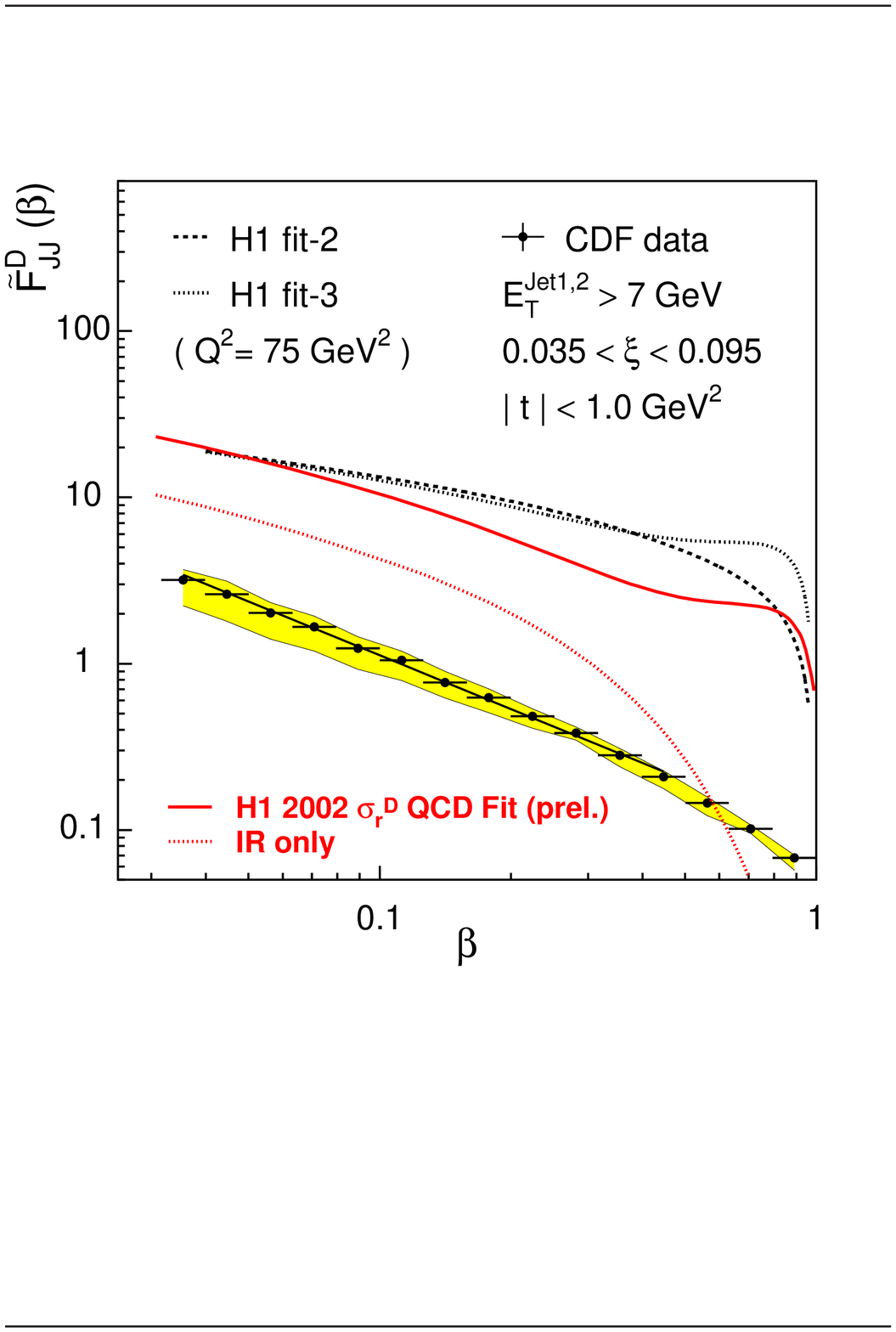,width=9cm,clip=true}
\vspace{8cm}
\caption{Comparison between the CDF measurement of diffractive structure
function (black points) with the expectation of the H1 QCD fits (red full line).}
\label{fig4}
\end{center}
\end{figure}

The second step is to check whether factorisation holds or not between Tevatron and
HERA data. The measurement of the diffractive structure function is possible
directly at the Tevatron. The CDF collaboration measured the ratio of dijet
events in single diffractive and non diffractive events, which is directly
proportional to the ratio of the diffractive to the ``standard" proton structure
functions $F_2$:
\begin{eqnarray}
R(x) = \frac{Rate^{SD}_{jj} (x)}{Rate^{ND}_{jj} (x)} \sim
\frac{F^{SD}_{jj} (x)}{F^{ND}_{jj} (x)}
\end{eqnarray}
The ``standard" proton structure function is known from the usual PDFs obtained
by the
CTEQ or MRST parametrisations. The comparison between the CDF measurement 
(black points, with systematics errors as shaded area) and the
expectation from the H1 QCD fits in full line is shown in 
Fig.~\ref{fig4}~\cite{cdffact}. 
We notice a discrepancy of a factor 8 to 10 between the data and the predictions from
the QCD fit, showing that factorisation does not hold. However, the difference
is compatible within systematic and statistical uncertainties
with a constant on a large part of the kinematical plane in
$\beta$, showing that the survival probability does not seem to be
$\beta$-dependent within experimental uncertainties. It would be interesting
to make these studies again in a wider kinematical domain both at the Tevatron and at
the LHC. The understanding of the survival probability and its dependence on the
kinematic variables is important to make precise predictions on inclusive diffraction
at the LHC.

\begin{figure}
\begin{center}
\epsfig{file=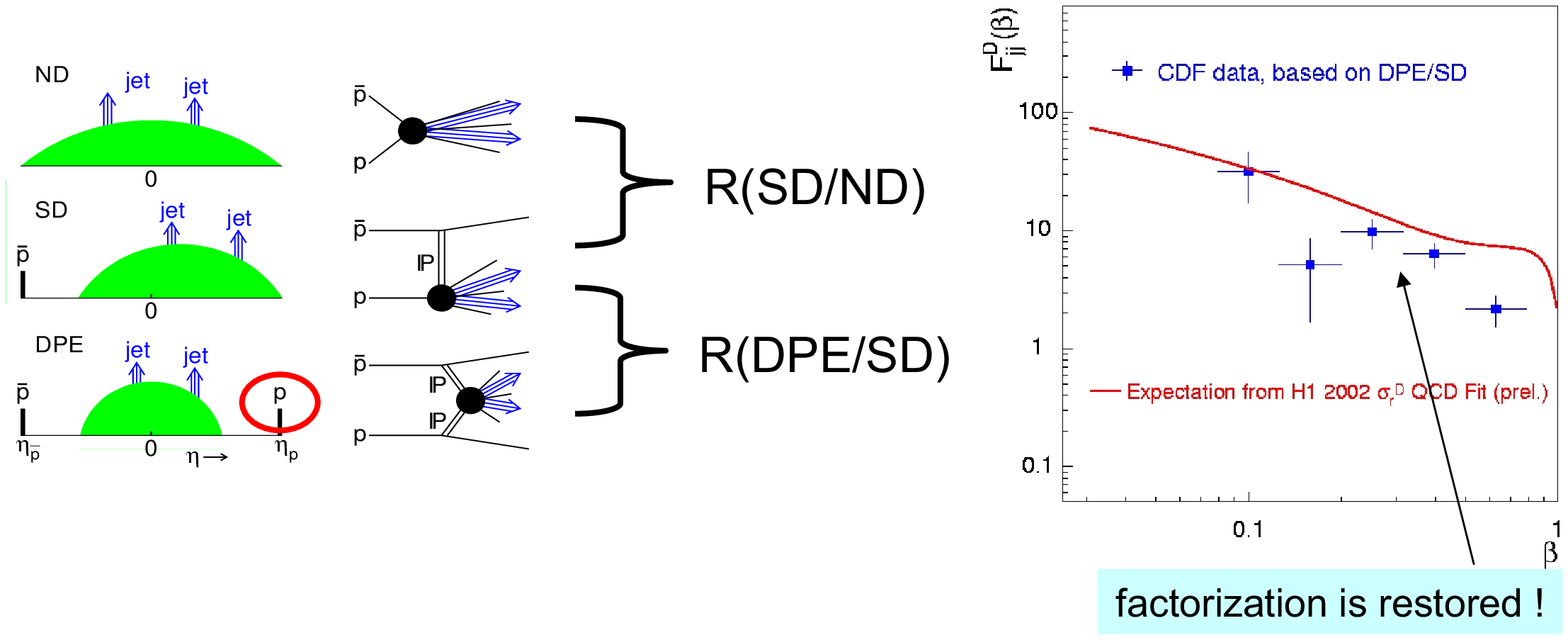,width=6cm,angle=270}
\caption{Restoration of factorisation for the ratio of double pomeron exchange
to single diffractive events (CDF Coll.).}
\label{fig5}
\end{center}
\end{figure}

The other interesting measurement which can be also performed at the Tevatron is
the test of factorisation between single diffraction and double pomeron
exchange. The results from the CDF collaboration are shown in 
Fig.~\ref{fig5}~\cite{cdffact}.
The left plot shows the definition of the two ratios while the right figure
shows the comparison between the ratio of double pomeron exchange to single
diffraction and the QCD predictions using HERA data in full line. Whereas
factorisation was not true for the ratio of single diffraction to non diffractive events,
factorisation holds for the ratio of double pomeron exchange to single
diffraction! In other words, the price to pay for one gap is the same as the
price to pay for two gaps. The survival probability, i.e. the probability not to emit an additional
soft gluon after the hard interaction needs to be applied only once to require
the existence of a diffractive event, but should not be applied again for double
pomeron exchange.

To summarize, factorisation does not hold between HERA and Tevatron as expected
because of the long term additional soft exchanges with respect to the the hard
interaction. However, experimentally, factorisation holds with CDF data themselves and also
between single diffraction and double pomeron exchange which means that the soft
exchanges do not depend on the hard scattering, which is somehow natural.

\subsection{Possibility of survival probablity measurements at D\O\ }
As we mentionned already, it is very important to test and understand the concept of
survival probability.
A new measurement can be performed at the Tevatron,
in the D\O\ experiment, which can be
decisive to test directly the concept of survival probability at the Tevatron, 
by looking at the azimuthal 
distributions of the outgoing proton and antiproton with respect to the 
beam direction~\cite{alexander}. 

\begin{figure}
\begin{center}
\epsfig{file=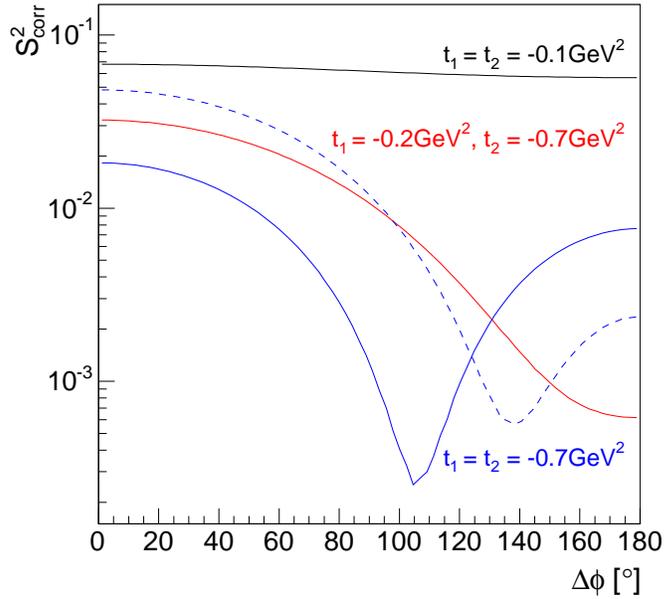,width=9cm}
\caption{$\Delta \Phi$ dependence of the survival probability for two different
models of survival probability where $\Delta \Phi$ is the difference in
azimuthal angle between the scattered $p$ and $\bar{p}$ in the final state, and
for three different values of $t$ (see text).}
\label{fig6}
\end{center}
\end{figure}

In Fig.~\ref{fig6}, we display the survival probability for three different
values of $t$ as a function of the difference in azimuthal angle between the
scattered $p$ and $\bar{p}$. The upper black curve represents the case where the
$t$ of the $p$ and $\bar{p}$ are similar and close to 0. In that case, only a
weak dependence on $\Delta \Phi$ is observed. The conclusion is different for
asymmetric cases or cases when $t$ is different from 0: Fig.~\ref{fig6}
also shows the result in full red line for the asymmetric case ($t_1=0.2$,
$t_2=0.7$ GeV$^2$), and in full and dashed blue lines for $t_1=t_2=0.7$ GeV$^2$
for two different models of survival probabilities. We notice that we get a very
strong $ \Delta \Phi$ dependence of more than one order of magnitude. 

The $\Delta \Phi$ dependence can
be tested directly using the roman pot detectors at D\O\ (dipole
and quadrupole detectors) and their possibility
to measure the azimuthal angles of the $p$ and $\bar{p}$~\cite{alexander}. 

The possible measurements can also be compared
to expectations using another kind of model to describe diffractive events,
namely soft colour interaction~\cite{sci}. This model assumes that diffraction
is not due to a colourless exchange at the hard vertex (called pomeron) but
rather to string rearrangement in the final state during hadronisation. In this
kind of model, there is a probability (to be determined by the experiment) that
there is no string connection, and so no colour exchange, between the partons
in the proton and the scattered quark produced during the hard interaction.
Since this model does not imply the existence of pomeron, there is no need of a
concept like survival probability, and no dependence on $\Delta \Phi$ of
diffractive cross sections. The proposed measurement would allow to distinguish
between these two drastically different models of diffraction.

\section{Diffractive exclusive event production}

\subsection{Interest of exclusive events}
\begin{figure}
\begin{center}
\epsfig{file=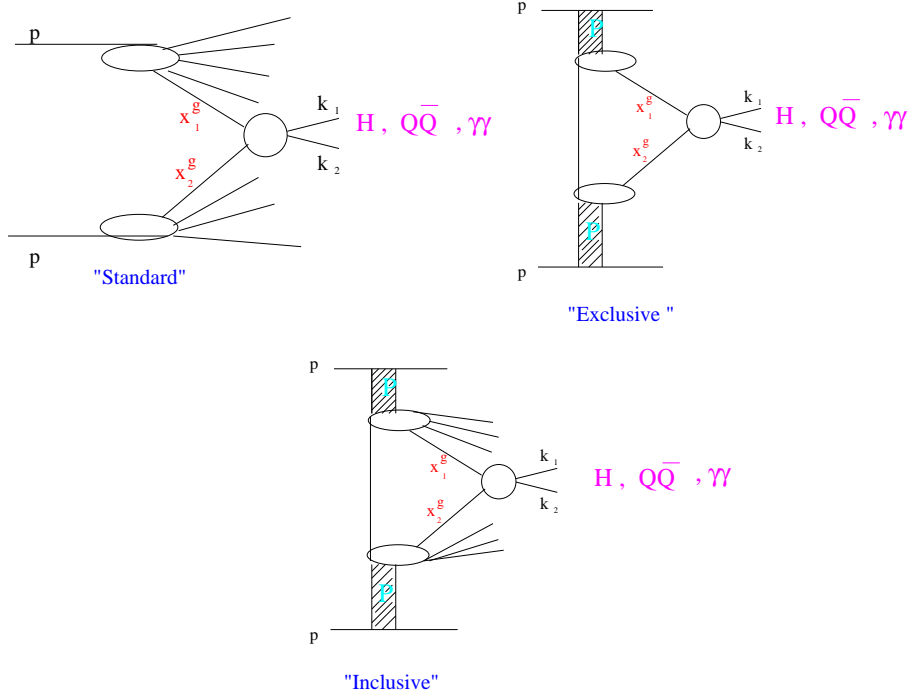,width=12cm}
\caption{Scheme of non diffractive, inclusive double pomeron exchange,
exclusive diffractive events at the Tevatron or the LHC.}
\label{fig7}
\end{center}
\end{figure}

A schematic view of non diffractive, inclusive double pomeron exchange,
exclusive diffractive events at the Tevatron or the LHC is displayed in
Fig.~\ref{fig7}.
The upper left plot shows the ``standard" non diffractive events
where the Higgs boson, the dijet or diphotons are produced directly by a
coupling to the proton and shows proton remnants. The bottom plot displays 
the standard diffractive double
pomeron exchange where the protons remain intact after interaction and the total
available energy is used to produce the heavy object (Higgs boson, dijets,
diphotons...) and the pomeron remnants. We have so far only discussed
this kind of events and their diffractive production using the
parton densities measured at HERA. There may be a third class of processes
displayed in the upper right figure, namely the exclusive diffractive
production. In this kind of events, the full energy is used to produce the heavy
object (Higgs boson, dijets, diphotons...) and no energy is lost in pomeron
remnants. There is an important kinematical consequence: the mass of the
produced object can be computed using roman pot detectors and tagged protons:

\begin{eqnarray}
M = \sqrt{\xi_1 \xi_2 S}.
\end{eqnarray} 
We see immediately the advantage of those processes: we can benefit from the
good roman pot resolution on $\xi$ to get a good resolution on mass. It is then
possible to measure the mass and the kinematical properties of the produced
object and use this information to increase the signal over background ratio by reducing the
mass window of measurement. It is thus important to know if this kind of events
exist or not. We will now describe in detail the search for exclusive events in the
different channels which is performed by the CDF and D0 collaborations at the Tevatron.
In the next section, we will also discuss the impact of the exclusive events on the
LHC physics potential.

\subsection{Search for exclusive events in $\chi_c$ production}
One way to look for exclusive events at the Tevatron is to search 
for the diffractive exclusive production of 
light particles like the $\chi$ mesons. This would give rise to high enough 
cross sections -- contrary to the diffractive exclusive production of heavy
mass objects such as Higgs bosons --- to check the dynamical 
mechanisms and the existence of exclusive events. 
Indeed, exclusive production of $\chi_{c}$ has been studied by the 
CDF collaboration~\cite{cdfchic} with an upper limit for the cross section 
of $\sigma_{exc}(p\bar{p} \rightarrow p+J/\psi
 + \gamma+\bar{p}) \sim 49 \pm 18 (stat)
\pm 39 (sys)\  $ pb where the $\chi_c$ decays into $J/\Psi$ and $\gamma$, the
$J/\Psi$ decaying itself into two muons. The experimental signature is thus two
muons in the final state and an isolated photon, which is a very clear signal.
Unfortunately, the cosmics contamination is difficult to compute and this is
why the CDF collaboration only quotes an upper limit on the $\chi_c$ production
cross section.
To know if the production is really
exclusive, it is important to study the tail of inclusive diffraction which is a
direct contamination of the exclusive signal. The tail of inclusive diffraction
corresponds to events which show very little energy in the forward direction, or in
other words where the pomeron remants carry very little energy. This is why these
events can be called quasi-exclusive.
In Ref.~\cite{chic}, we found that the contamination of
inclusive events into the signal region depends strongly on the
assumptions on the gluon distribution in the pomeron at high $\beta$ --- which
is very badly known as we mentioned in a previous section. 
Therefore, this channel is unfortunately not
conclusive concerning the existence of exclusive events.

\subsection{Search for exclusive events in the diphoton channel}
The CDF collaboration also looked for the exclusive production of dilepton and
diphoton~\cite{cdfgamma}. 
Contrary to diphotons, dileptons cannot be produced exclusively via pomeron exchanges since
$g g \rightarrow \gamma \gamma$ is possible, but $g g \rightarrow l^+ l^-$ 
directly is impossible. However, dileptons can be produced via QED processes, and
the cross section is perfectly known. The CDF dilepton measurement is $\sigma = 1.6
^{+0.5}_{-0.3} (stat) \pm 0.3 (syst)$ pb which is found to be in good agreement
with QED predictions and shows that the acceptance, efficiencies of the detector
are well understood. 3 exclusive diphoton events have been observed by the CDF
collaboration leading to a cross section of
$\sigma = 0.14
^{+0.14}_{-0.04} (stat) \pm 0.03 (syst)$ pb compatible with the expectations
for exclusive diphoton production at the Tevatron. Unfortunately, the number of events
is very small and the cosmics contamination uncertain. The conclusions about the
existence of exclusive events are thus uncertain. This channel will be however very
important for the LHC where the expected exclusive cross section is much higher.

\subsection{Search for exclusive events using the dijet mass fraction at the Tevatron}

The CDF collaboration measured the so-called dijet mass fraction
in dijet events --- the ratio of the mass carried by the two jets produced in the event 
divided by the
total diffractive mass --- when the antiproton is tagged in the roman pot
detectors and when there is a rapidity gap on the proton side to ensure that the
event corresponds to a double pomeron exchange. 
The CDF collaboration measured this
quantity for different jet $p_T$ cuts~\cite{cdfrjj}. We compare this measurement
to the expectation coming from the structure of the pomeron coming from HERA.
For this sake, one takes the gluon and quark densities in the pomeron measured at HERA
as described in Ref.~\cite{loehr,us} and the factorisation breaking between
HERA and the Tevatron is assumed to come only through the gap survival probability
(0.1 at the Tevatron). 

\begin{figure}
\centerline{\includegraphics[width=0.7\columnwidth]{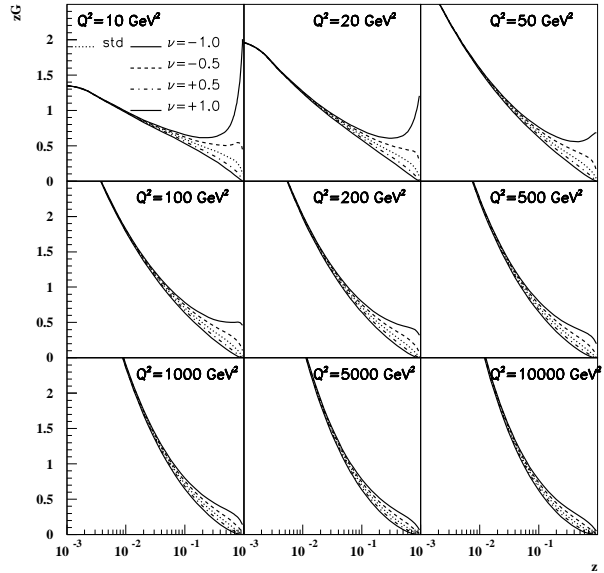}}
\caption{Uncertainty of the gluon density at high $\beta$ (here $\beta\equiv z$).
The gluon density is multiplied by the factor $(1-\beta)^{\nu}$ where $\nu$=-1.,
-0.5, 0.5, 1. The default value $\nu =0$ is the gluon density in the pomeron
determined directly by a fit to the H1 $F_2^D$ data with an uncertainty
of about 0.5.}
\label{gluon}
\end{figure}

In Fig.~\ref{gluon}, we recall the gluon and quark densities in
the pomeron measured at HERA. The gluon density at high $\beta$
is not well constrained from the QCD fits performed at HERA. To study this uncertainty, we multiply the gluon distribution by the
factor $(1 - \beta)^{\nu}$ as shown in Fig.~\ref{gluon}. The $\nu$ parameter
varies between -1 and 1. For $\nu=-$1, the gluon density in the pomeron is
enhanced at high $\beta$ whereas it is damped when $\nu=$1.
QCD fits to the H1 data lead to 
the uncertainty on the $\nu$ parameter $\nu=0.0\pm0.5$~\cite{us}.

\begin{figure}
\begin{tabular}{cc}
\hspace{-1cm}
\epsfig{figure=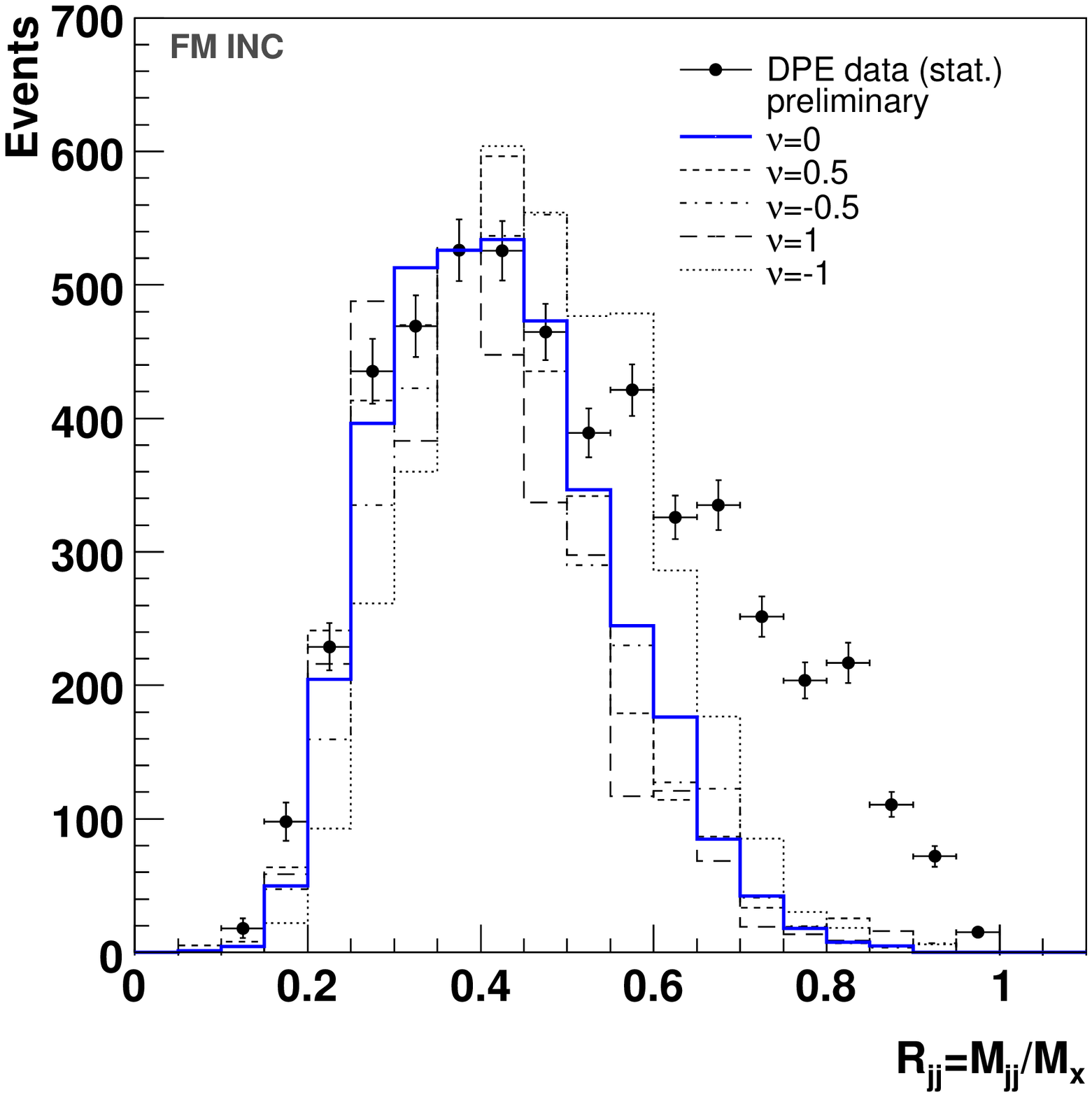,height=2.5in}  &
\epsfig{figure=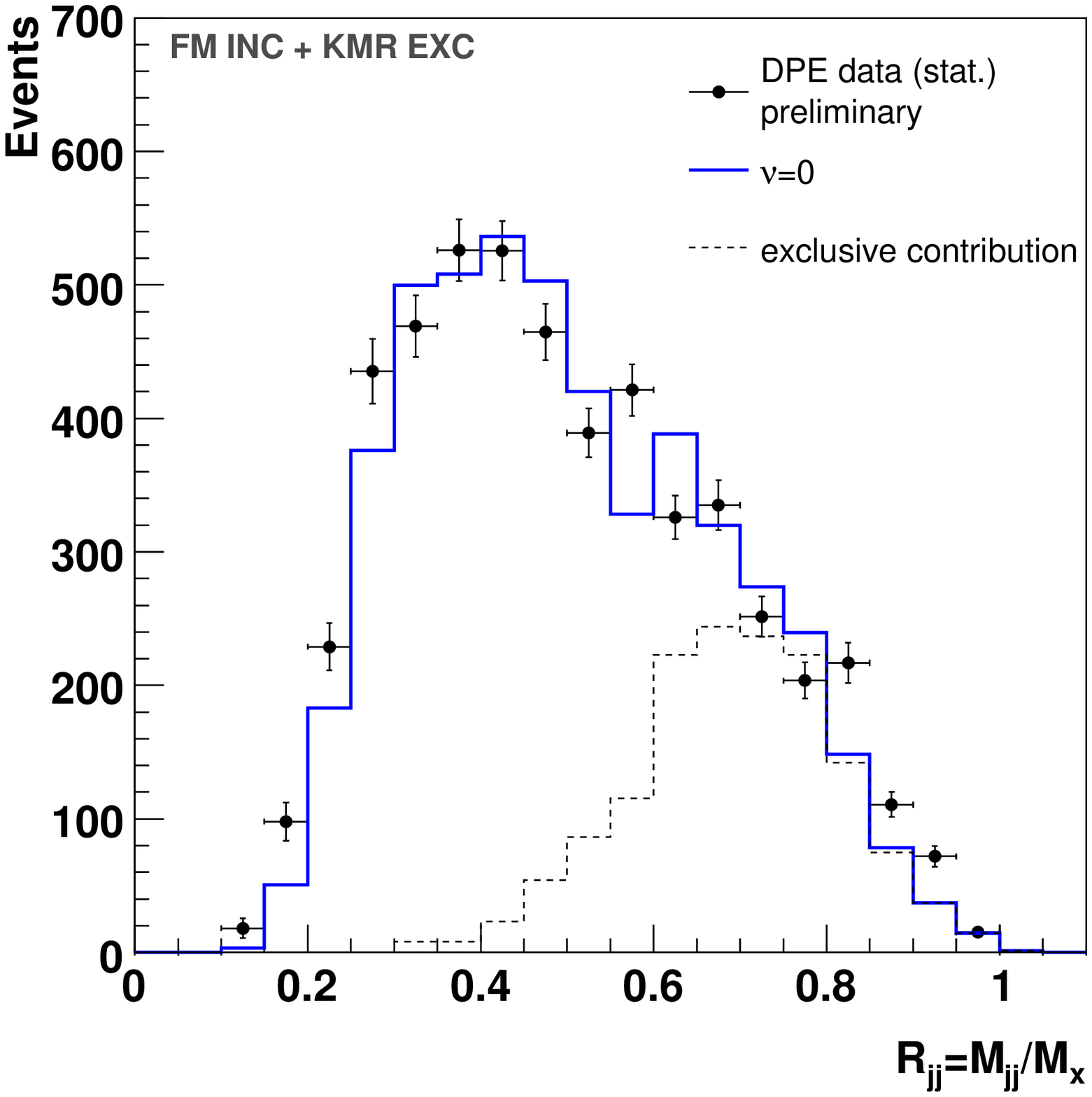,height=2.5in} \\
\end{tabular}
\caption{Left: Dijet mass fraction measured by the CDF collaboration compared to the
prediction from inclusive diffraction based on the parton densities
in the pomeron measured at HERA. The gluon
density in the pomeron at high $\beta$ was modified by varying the parameter
$\nu$.
Right: Dijet mass fraction measured by the CDF collaboration compared to the
prediction adding the contributions from inclusive and exclusive diffraction.}
\label{compare2}
\end{figure}

The comparison between the CDF data for a jet $p_T$ cut of 10 GeV as an
example and the predictions from inclusive diffraction is given in 
Fig.~\ref{compare2}, left. 
We also give
in the same figure the effects of changing the gluon density at high $\beta$ (by
changing the value of the $\nu$ parameter) and we note that inclusive
diffraction is not able to describe the CDF data at high dijet mass fraction,
even after increasing the gluon density in the pomeron at high $\beta$ (multiplying
it by $1/(1-\beta)$),
where exclusive events are expected to appear~\cite{oldab}. The conclusion
remains unchanged when jets with $p_T>25$ GeV are considered~\cite{oldab}.

Adding exclusive events to the distribution of the dijet mass fraction leads to
a good description of data~\cite{oldab} as shown in Fig.~\ref{compare2}, right, 
where we superimpose the
predictions from inclusive and exclusive diffraction. This study does not prove
that exclusive events exist but shows that some additional component with respect to
inclusive diffraction is needed to explain CDF data. Adding exclusive
diffraction allows to explain the CDF measurement. To be sure of the existence
of exclusive events, the observation will have to be done in different channels
and the different cross sections to be compared with theoretical expectations.
In Ref.~\cite{oldab}, the CDF data were also compared to the soft colour
interaction models~\cite{sci}. While the need for exclusive events is less
obvious for this model, especially at high jet $p_T$, the jet rapidity
distribution measured by the CDF collaboration is badly reproduced. This is due
to the fact that, in the SCI model, there is a large difference between
requesting an intact proton in the final state and a rapidity gap.
 
\begin{figure}
\centerline{\includegraphics[width=0.7\columnwidth]{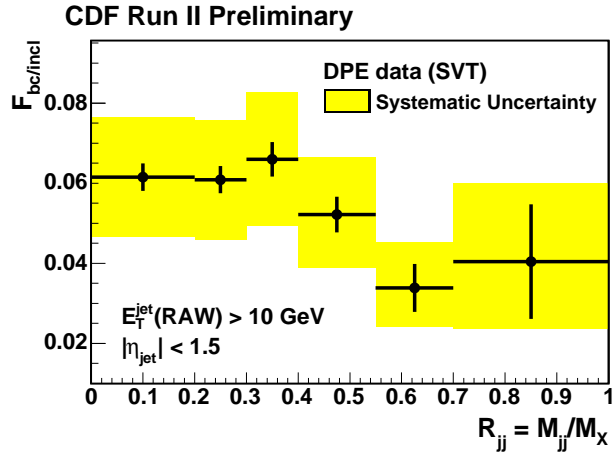}}
\caption{Ratio of $b/c$ jets to inclusive jets in double pomeron exchange events
as a function of the dijet mass fraction.}
\label{bjet}
\end{figure}

Another interesting observable in the dijet channel is to look at the fraction
of $b$ jets as a function of the dijet mass fraction. In exclusive events, the
$b$ jets are suppressed because of the $J_Z=0$ selection rule~\cite{ushiggs}, and it is
expected that the fraction of $b$ jets in the diffractive dijet sample
diminishes as a function of the dijet mass fraction. The results from the CDF
collaboration are given
in Fig.~\ref{bjet}~\cite{cdfrjj}. We see a tendency of the $b/c$ jet fraction in
data to go down as a function of the dijet mass ratio but the statistics is
still low.

Another way to look for exclusive events is to study the
correlation between the gap size measured in both $p$ and $\bar{p}$ directions
and the value of $\log 1/\xi$ measured using roman pot detectors~\cite{tev4lhc}. 
The gap size between the
pomeron remnant and the protons detected in roman pot detector 
is of the order of 
$\log 1/\xi$ for usual diffractive events while
exclusive events show a larger rapidity gap since the gap
occurs between the jets and the proton detected in roman
pot detectors (in other words, there is no pomeron remnant). A way to see
exclusive events would be for instance to look for diffractive events in the
two jet event sample in the central part of the calorimeter with a
a dijet mass above a high enough mass threshold $M_{threshold}$
so that the size of the expected
gap is small ($\sqrt{\xi_1 \xi_2} > M_{threshold}/\sqrt{s}$ and $gap~size~\sim
\log(1/\xi)$). If some events are found with a much larger rapidity gap, they
should be exclusive since the gap is between the jet and the proton and not
between the pomeron remnant and the proton.

\subsection{Search for exclusive events at the LHC}
The search for exclusive events at the LHC can be performed in the same channels
as the ones used at the Tevatron. In addition, some other possibilities
benefitting from the high luminosity of the LHC appear. One of the cleanest ways
to show the existence of exclusive events would be to measure the dilepton and
diphoton cross section ratios as a function of the 
dilepton/diphoton mass~\cite{ushiggs,tev4lhc}. If
exclusive events exist, this distribution should show a bump towards high values
of the dilepton/diphoton mass since it is possible to produce exclusively
diphotons but not dileptons at leading order as we mentionned in the previous
paragraph. 

The search for exclusive events at the LHC will also require a precise analysis
and measurement of inclusive diffractive cross sections and in particular the
tails at high $\beta$ since it is a direct
background to exclusive event production. It will be also useful to measure directly
the exclusive jet production cross section as a function of jet $p_T$ as an example
and compare the evolution to the models. This will allow to know precisely the
background especially to Higgs searches which we will discuss in the following.

\section{Diffraction at the LHC}
In this section, we will describe briefly some projects concerning diffraction
at the LHC. We will put slightly more emphasis on the diffractive production
of heavy objects such as Higgs bosons, top  or stop pairs, $WW$ events, etc...

\subsection{Diffractive event selection at the LHC}
The LHC with a center-of-mass energy of 14 TeV will allow us to access a completely
new kinematical domain in diffraction. So far, three experiments, namely ATLAS and
CMS-TOTEM have shown interests in diffractive measurements.
The diffractive event selection at the LHC will be the same as at the Tevatron.
However, the rapidity gap selection will no longer be possible at high
luminosity since up to 35 interactions per bunch crossing  are expected to occur
and soft pile-up events will kill the gaps produced by the hard interaction.
Proton tagging will thus be the only possibility to detect diffractive events at
high luminosity. Let us note that this is not straightforward:
we need to make sure that the diffracted protons come from the hard
interaction and not from the soft pile up events. The idea we will develop in
the following is to measure precisely the time of arrival of the diffracted
protons in the forward detectors, and thus know if the protons come from the
vertex of the hard interaction.

\subsection{Measurements at the LHC using a high $\beta^*$ lattice in ATLAS-ALFA
and TOTEM}
\begin{figure}
\begin{center}
\epsfig{figure=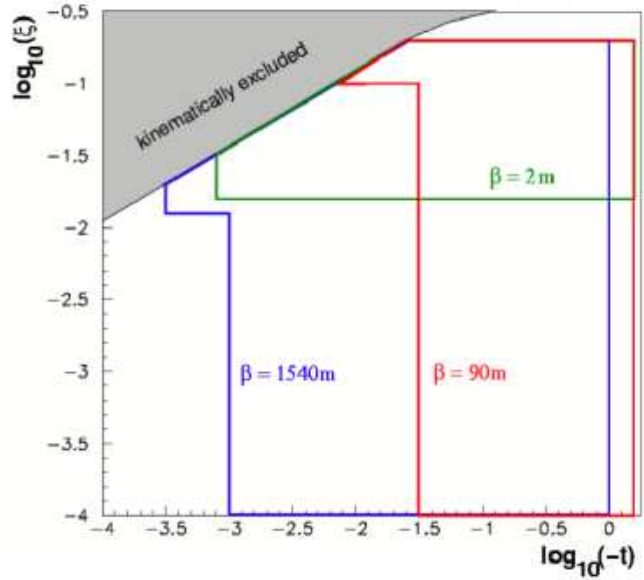,height=3.1in}  
\caption{\label{totem1} TOTEM acceptance for different lattices. 
}
\end{center}
\end{figure}

\begin{figure}
\epsfig{figure=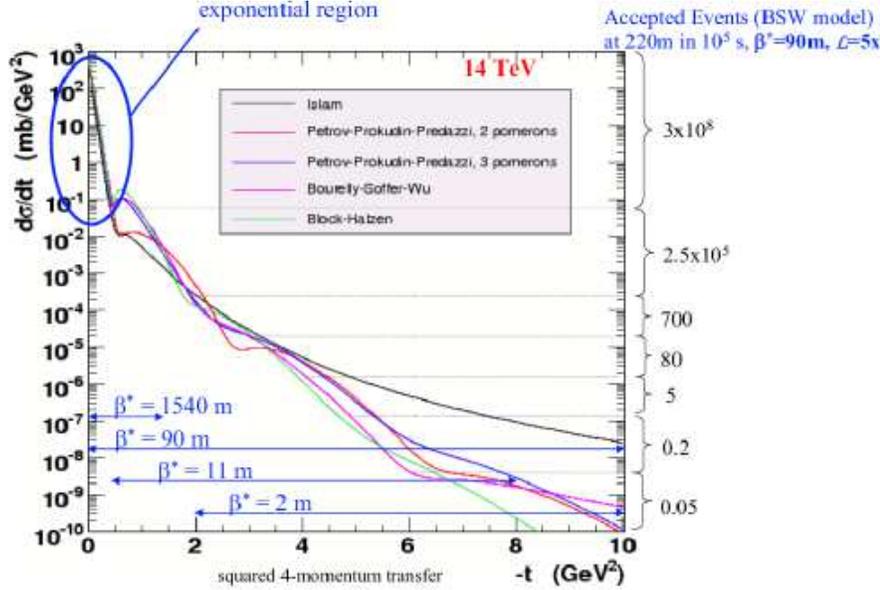,height=3.1in} 
\caption{\label{totem2} 
Measurement of the elastic section and domain accessible for different
$\beta^*$.}
\end{figure}

Measurements of total cross section and luminosity are foreseen in the 
ATLAS-ALFA~\cite{atlaslumi} and TOTEM~\cite{totem} experiments, and roman pots are
installed at 147 and 220 m in TOTEM and 240 m in ATLAS. These measurements will
require a special injection lattice of the LHC at low luminosity since they require the
roman pot detectors to be moved very close to the beam. The acceptance of the
TOTEM detectors for different injection lattices is given in Fig.~\ref{totem1},
and the possibilities to measure the $t$-dependence of
the total cross section using the different injection lattices in
Fig.~\ref{totem2}.
We notice that high $\beta^*$ lattices are needed if one wants to access the
low-$t$ measurement of the total cross section.

The
measurement of the total cross section to be performed by the TOTEM
collaboration~\cite{totem} is shown in Fig.
22. We notice that there is a large uncertainty on prediction of the total cross
section at the LHC energy in particular due to the discrepancy between the two
Tevatron measurements. The inelastic $p \bar{p}$ cross section was measured at
a center-of-mass energy of 1.8 TeV at the Tevatron by the E710, E811 and CDF
collaborations which lead to the following respective results: $56.6 \pm 2.2$ mb,
$56.5 \pm 1.2$ mb and $61.7 \pm 1.4$ mb~\cite{tevtotal}. While the E710 and E811 experiments agree (E811 is
basically the follow up of E710), the E811 and CDF measurements disagree by
9.2\%, and the reason is unclear~\cite{tevtotal}.
The measurement of TOTEM will be of special
interest to solve that ambiguity as well.

\begin{figure}
\begin{center}
\epsfig{file=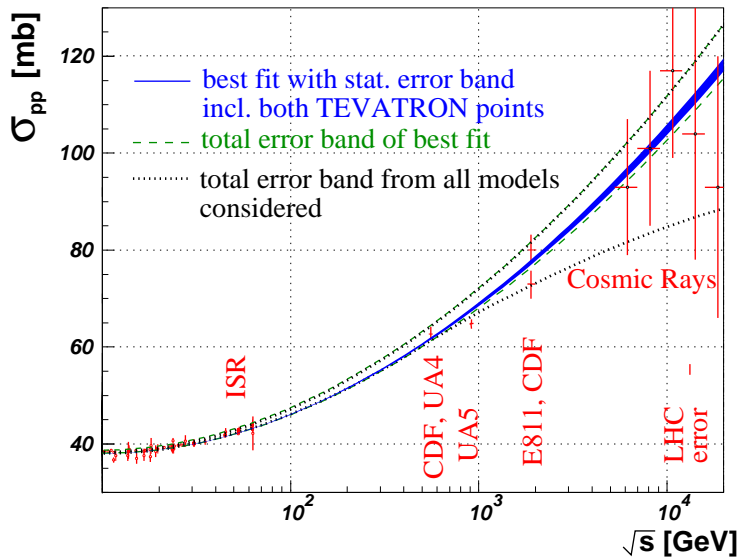,width=10cm}
\caption{Measurement of the total cross section.}
\label{fig18}
\end{center}
\end{figure} 

The ATLAS collaboration prefers to measure the elastic scattering in the Coulomb
region~\cite{atlaslumi}, typically at very low $t$ ($|t| \sim 6.5~10^{-4}$ GeV$^2$). When $t$ is
close to 0, the $t$ dependence of the elastic cross section reads:
\begin{eqnarray}
\frac{dN}{dt} (t \rightarrow 0) = L \pi \left( \frac{-2 \alpha}{|t|} +
\frac{\sigma_{tot}}{4 \pi} (i+\rho)e^{-b|t|/2} \right)^2.
\end{eqnarray}
From a fit to the data in the Coulomb region, it is possible to determine
directly the total cross section $\sigma_{tot}$, the $\rho$ and $b$ parameters
as well as the absolute luminosity $L$. This measurement requires to go down to
$t \sim 6.5~10^{-4}$ GeV$^2$, or $\theta \sim 3.5~\mu$rad (to reach the
kinematical domain where the strong
amplitude equals the electromagnetic one). The UA4 collaboration already
performed such a measurement at the SPS and reached a precision of the order of
3\%. However, the UA4 experiment~\cite{ua4} needed to perform a measurement down to 120 $\mu$rad whereas the
ATLAS collaboration needs to go down to 3.5 $\mu$rad, which is very challenging.
This measurement requires a special high $\beta^*$ lattice, the detectors to be
installed 1.5 mm from the LHC beam, a spatial resolution of these detectors well
below 100 $\mu$m and no significant dead edge on the detector (less than 100
$\mu$m). 

The solution to perform this measurement is to install two sets of 
roman pot detectors on each side of ATLAS located at about 240 m from the
interaction point, which can go close to the beam when the beam is stable.
Each roman pot is itself made of two detectors in the vertical direction.
The detector installed in the roman pot is made of 20 $\times$ 64 square
0.5 $\times$ 0.5 mm$^2$ scintillating fibers on ceramic substrate read out by
24 Multianode photomultipliers with 64 channels. The detector follows a U/V
geometry with 45 degree stereo layers, 64 fibers per plane in a module, 10
double sided modules per pot. The up and down detectors overlap for relative
alignment purposes. 

To check the accuracy of the measurement within ATLAS, a
full simulation of elastic events was performed for two values of $t$: 
$t=7.~10^{-4}$~GeV$^2$ and $t=10^{-5}$~GeV$^2$. A fit to $dN/dt$ using 10
million events leads to a measurement of luminosity and the total cross section
with a statistical precision of 1.5\% and 0.74\% respectively~\cite{atlaslumi}. 

Once the absolute
luminosity and the total cross section are known using these methods, the
relative luminosity measurement as a function of time will be performed in ATLAS
using the LUCID detector (Luminosity measurement Using Cerenkov Integrating
Detectors)~\cite{atlaslumi}. The front face of the LUCID detector is located about 17 meters from
the ATLAS interaction point and covers a domain in rapidity of 5.4$<|\eta|<$6.1.
The principle of the LUCID detector is quite simple. 168 Aluminium tubes are
filled with $C_4F_{10}$ or isobutane at 1 or 2 bar pressure. Winston cones at
the end of each tube bring the Cerenkov light onto quartz fibers. It is thus
possible to measure the number of particles which are produced in the very
forward region which is directly related to the instantaneous luminosity. The
LUCID detector is mainly sensitive to primary particles only: much more light
comes from primary particles than from secondaries or soft particles. The time
resolution is about 140 ps which allows to determinate the luminosity bunch by
bunch at the LHC. The detector allows to obtain a linear relationship between
luminosity and the number of tracks counted in the detector which leads to an
easy measurement of luminosity.

\begin{figure}
\begin{center}
\begin{tabular}{cc}
\hspace{-1cm}
\epsfig{file=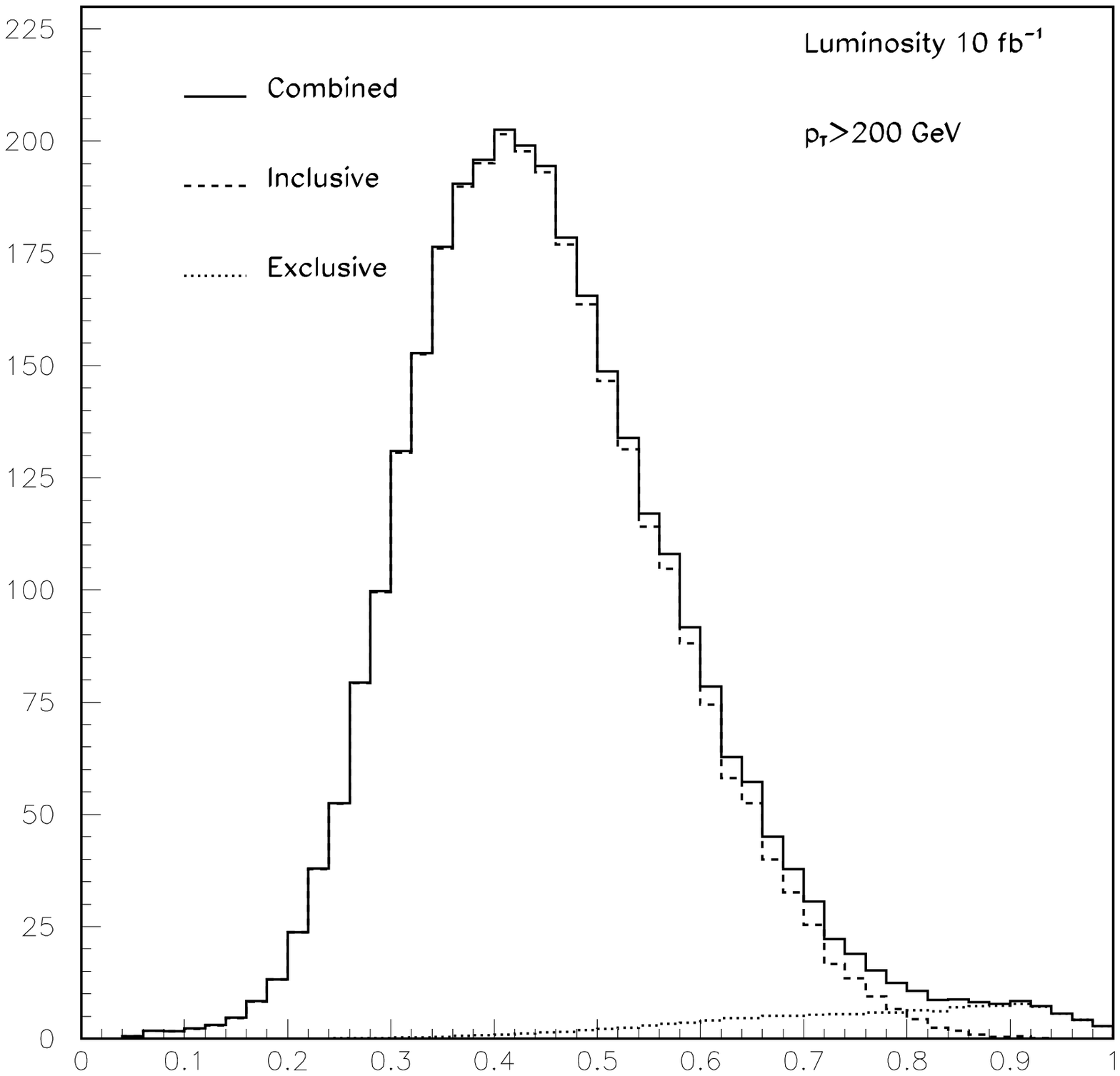,width=6.5cm} &
\epsfig{file=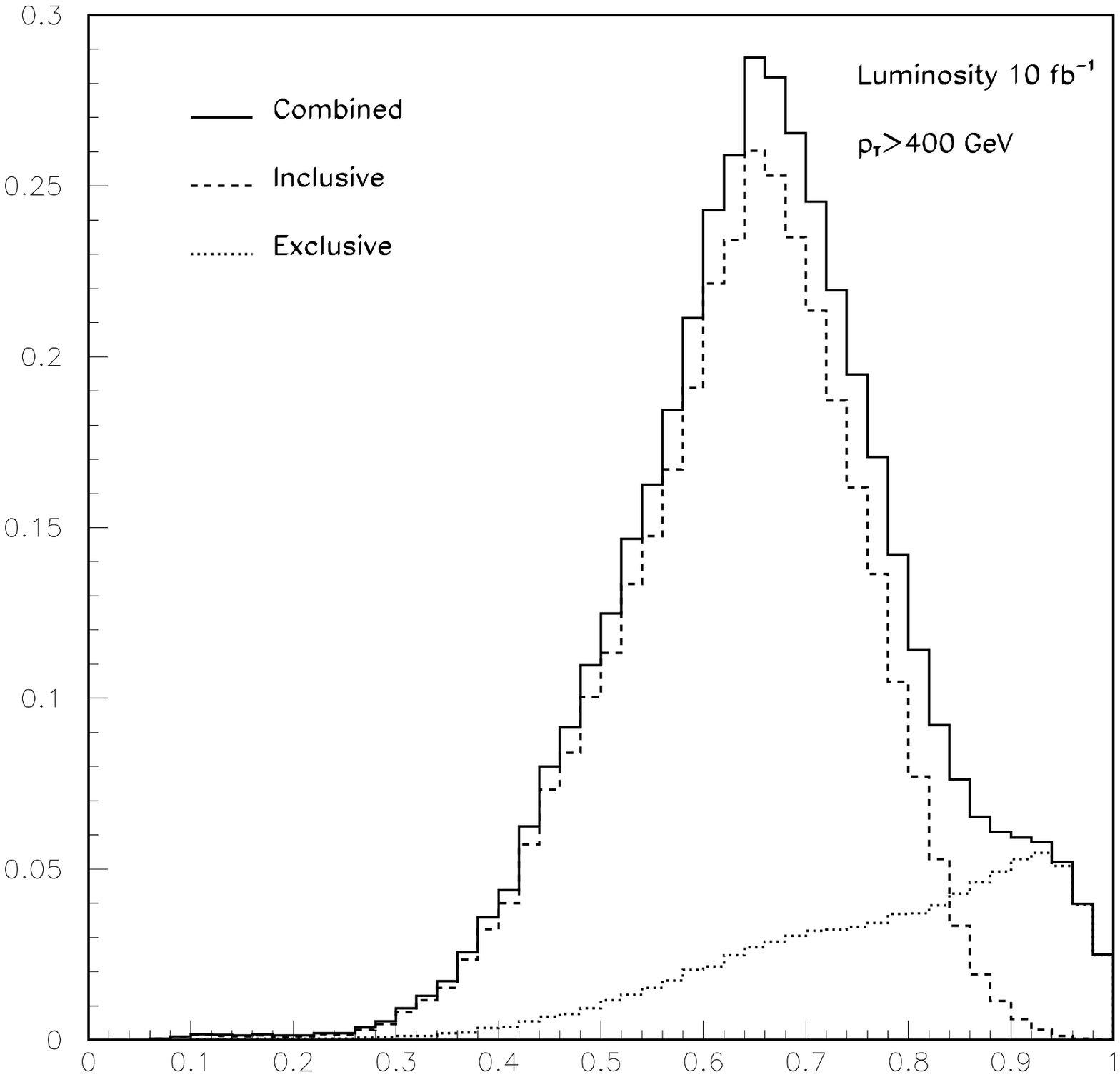,width=6.5cm} 
\end{tabular}
\caption{Dijet mass fraction at the LHC for jets $p_T>200\,\mathrm{GeV}$ and 
$p_T>400\,\mathrm{GeV}$ showing the contribution of both inclusive and exclusive
diffraction.}
\label{FigDMFexcLHC}
\end{center}
\end{figure}

\begin{figure}
\begin{center}
\epsfig{file=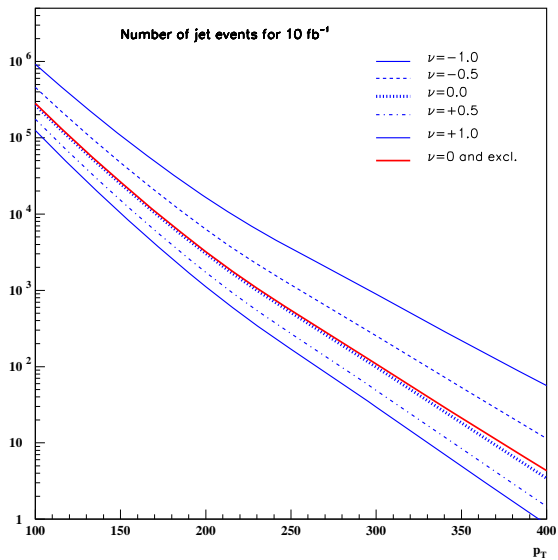,width=8.5cm} 
\caption{Number of DPE events at the LHC
as a function of minimal transverse momentum $p^{min}_T$ of the two leading jets
for different values of the $\nu$ parameter (describing the gluon in the
pomeron at high $\beta$). The effect of adding the exclusive contribution when
$\nu=0$ is also shown and is quite small with respect to the effects of the
different $\nu$ values. 
}
\label{FigNDPEptLHC}
\end{center}
\end{figure}

\begin{figure}
\begin{center}
\epsfig{file=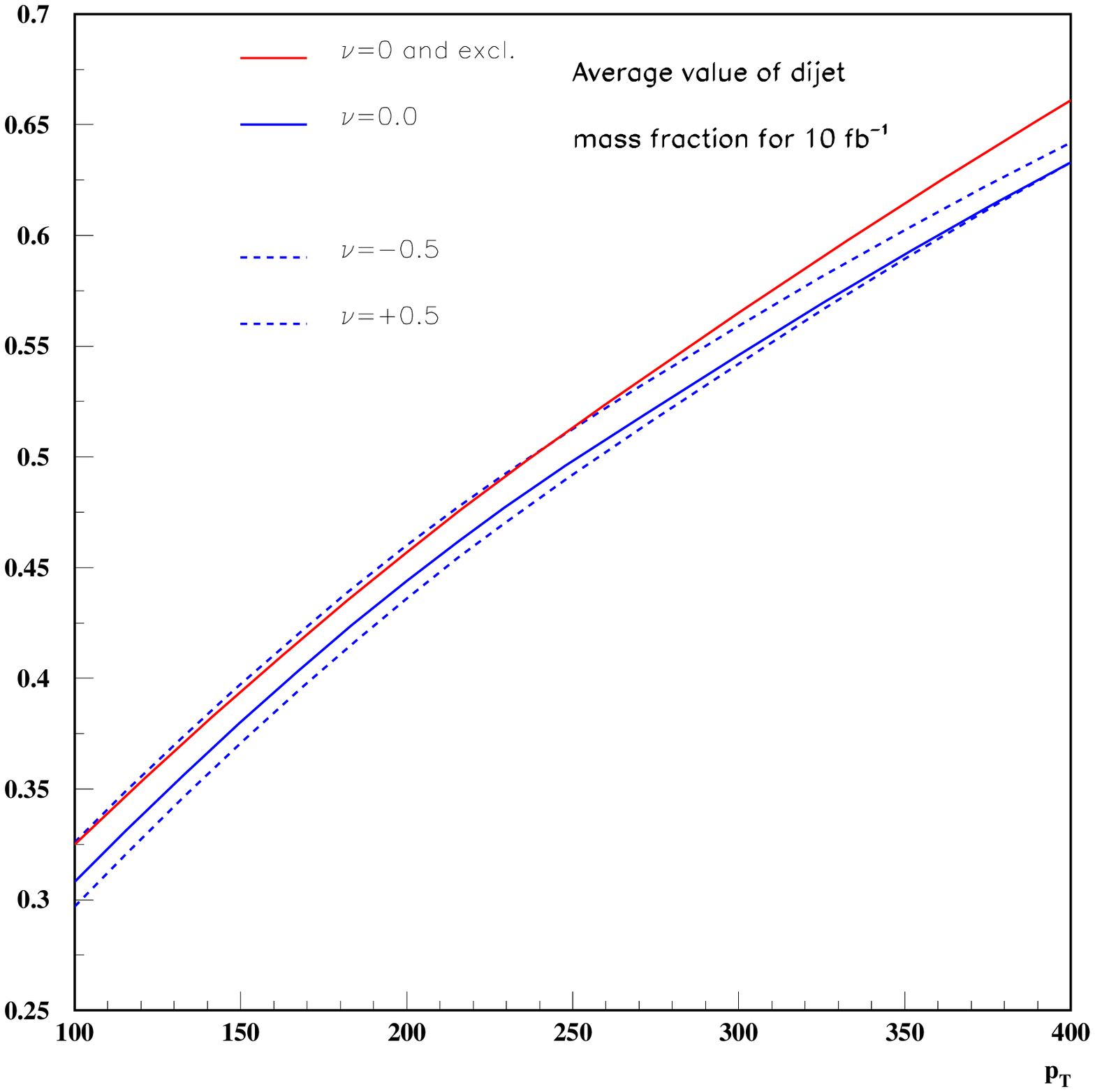,width=8.5cm}
\caption{Average value of the dijet mass fraction
as a function of minimal transverse momentum $p^{min}_T$ of the leading jets. 
Exclusive contribution and different values of $\nu$ 
are shown and its effect is clearly visible, especially for high jet transverse
momentum.}
\label{FigNDPEptLHCb}
\end{center}
\end{figure}

\subsection{Hard inclusive and exclusive diffraction at the LHC}
In this section, we would like to discuss first how we can measure the gluon density
in the pomeron, especially at high $\beta$ since the gluon in this kinematical 
domain shows large uncertainties and this is where the exclusive contributions
should show up. To take into account the high-$\beta$
uncertainties of the gluon distribution, we chose to multiply the gluon density
in the pomeron measured at HERA by a factor $(1-\beta)^{\nu}$  where $\nu$
varies between -1.0 and 1.0 as we already mentioned in a previous
section (see Fig.~\ref{gluon}).

The dijet mass fraction as a function of different jet $p_T$ is visible in 
Fig.~\ref{FigDMFexcLHC} after a simulation of the ATLAS/CMS detectors. The
exclusive contribution manifests itself as an increase in the tail of the 
distribution which can be seen for $200\,\mathrm{GeV}$ 
jets (left) and $400\,\mathrm{GeV}$ jets (right) respectively~\cite{oldab}. Exclusive production slowly 
turns on with the increase of the jet $p_T$ which is demonstrated in Fig. 
\ref{FigNDPEptLHC}, where the number of expected double pomeron exchange events 
for different values of $\nu$ with and without the exclusive contribution
is shown. However, 
with respect to the uncertainty on the gluon density this appearance is 
almost negligible. In that sense, it is possible to use the diffractive dijet
cross section measurement as a function of jet transverse momentum to measure
more precisely the gluon density in the pomeron at high $\beta$.
In Fig.~\ref{FigNDPEptLHCb}, we give the average value of the dijet mass
fraction for different values of the minimal jet transverse momentum and with and
without the exclusive contribution, and we see that exclusive events have the
tendency to increase sensibly the average value of the dijet mass fraction.
One can use the average position
of the dijet mass fraction as a function of the minimal jet transverse momentum $p_T^{min}$
to study the presence of the exclusive contribution, once the gluon density
in the pomeron is better known.
This is true especially for high $p_T$ jets.

The exclusive production at the LHC plays a minor role for low $p_T$ jets. Therefore, measurements e.g for $p_T<200\,\mathrm{GeV}$ where the inclusive production is dominant could be used 
to constrain the gluon density in the pomeron. Afterwards, one can look in the high $p_T$ jet region to extract the exclusive 
contribution from the tail of the dijet mass fraction. 

Other measurements already mentionned such as the diphoton, dilepton cross
section ratio as a function of the dijet mass, the $b$ jet, $W$ and
$Z$  cross section measurements will be also quite important at the LHC.

\subsection{Exclusive Higgs production at the LHC}

\begin{figure}
\begin{center}
\epsfig{file=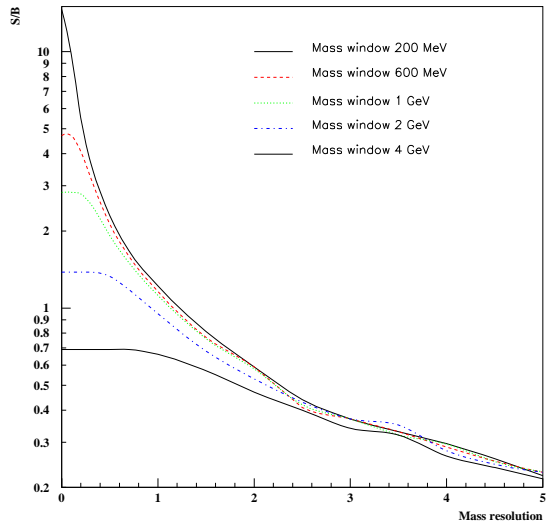,width=8cm} 
\caption{Standard Model Higgs boson signal to background ratio as a function 
of the resolution on the missing mass, in GeV. This figure assumes a Higgs
boson mass of 120 GeV.}
\label{fig20}
\end{center}
\end{figure}

\begin{figure}
\begin{center}
\epsfig{file=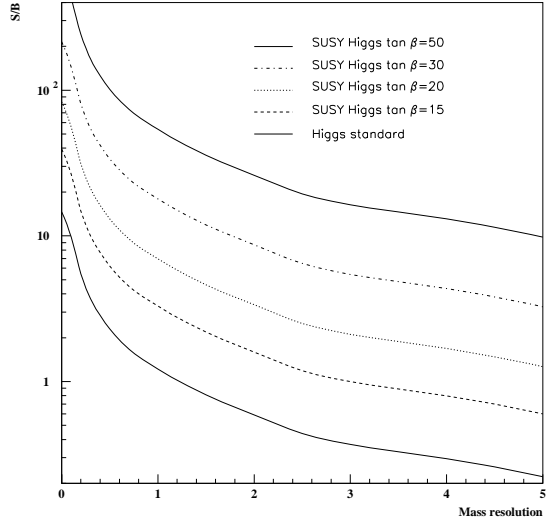,width=8cm}
\caption{SUSY Higgs boson signal to background ratio as a function 
of the resolution on
the missing mass, in GeV. This figure assumes a Higgs
boson mass of 120 GeV.}
\label{fig20b}
\end{center}
\end{figure}

As we already mentionned in one of the previous sections, one special interest of
diffractive events at the LHC is related to the existence of exclusive events
and the search for Higgs bosons at low mass in the diffractive mode.
So far, two projects are being discussed at the LHC: the installation of roman
pot detectors at 220 m in ATLAS~\cite{afp}, and at 420 m for the ATLAS and CMS
collaborations~\cite{fp420}. 

The results discussed in this section rely on the DPEMC Monte Carlo to produce 
Higgs bosons exclusively~\cite{ushiggs, dpemc} and a fast simulation of a
typical LHC detector (ATLAS or CMS). 
Results are given in Fig.~\ref{fig20} for a Higgs mass of 120 GeV, 
in terms of the signal to background 
ratio S/B, as a function of the Higgs boson mass resolution.
Let us notice that the background is mainly due to the exclusive $b \bar{b}$
production. However the tail of the inclusive $b \bar{b}$ production can also be
a relevant contribution and this is related to the high $\beta$ gluon
density which is badly known at present. The expected number of events after all
cuts is expected to be of the order of 5 for a luminosity of 60 fb$^{-1}$.
In order to obtain a S/B of 1, a mass resolution of about
1.2 GeV is needed which
is technically feasible, the inconvenient being the small number of events.

The diffractive SUSY Higgs boson production cross section is noticeably enhanced 
at high values of $\tan \beta$ and since we look for Higgs decaying into $b
\bar{b}$, it is possible to benefit directly from the enhancement of the cross
section contrary to the non diffractive case. A signal-over-background up to a
factor 50 can be reached for 100 fb$^{-1}$ for $\tan \beta \sim 50$
\cite{lavignac} (see Fig.~\ref{fig20b}). 

\begin{figure}
\begin{center}
\centerline{\psfig{figure=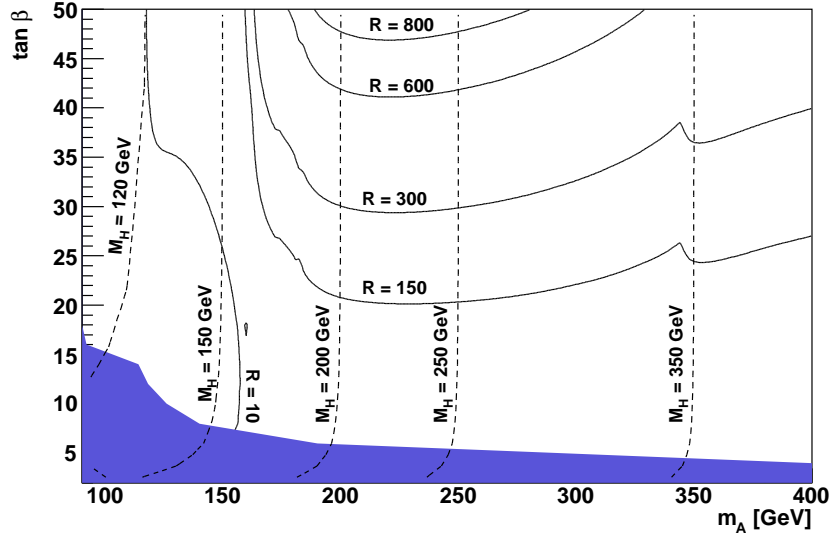,height=3.1in}}
\caption{Ratio $R$ at generator level between the number of diffractive Higgs
events in MSSM to SM in the ($\tan \beta$, $M_A$) plane. The lines of
constant Higgs boson mass are also indicated in dashed line.}
\label{marek}
\end{center}
\end{figure}

\begin{figure}
\begin{center}
\begin{tabular}{cc}
\hspace{-1cm}
\epsfig{figure=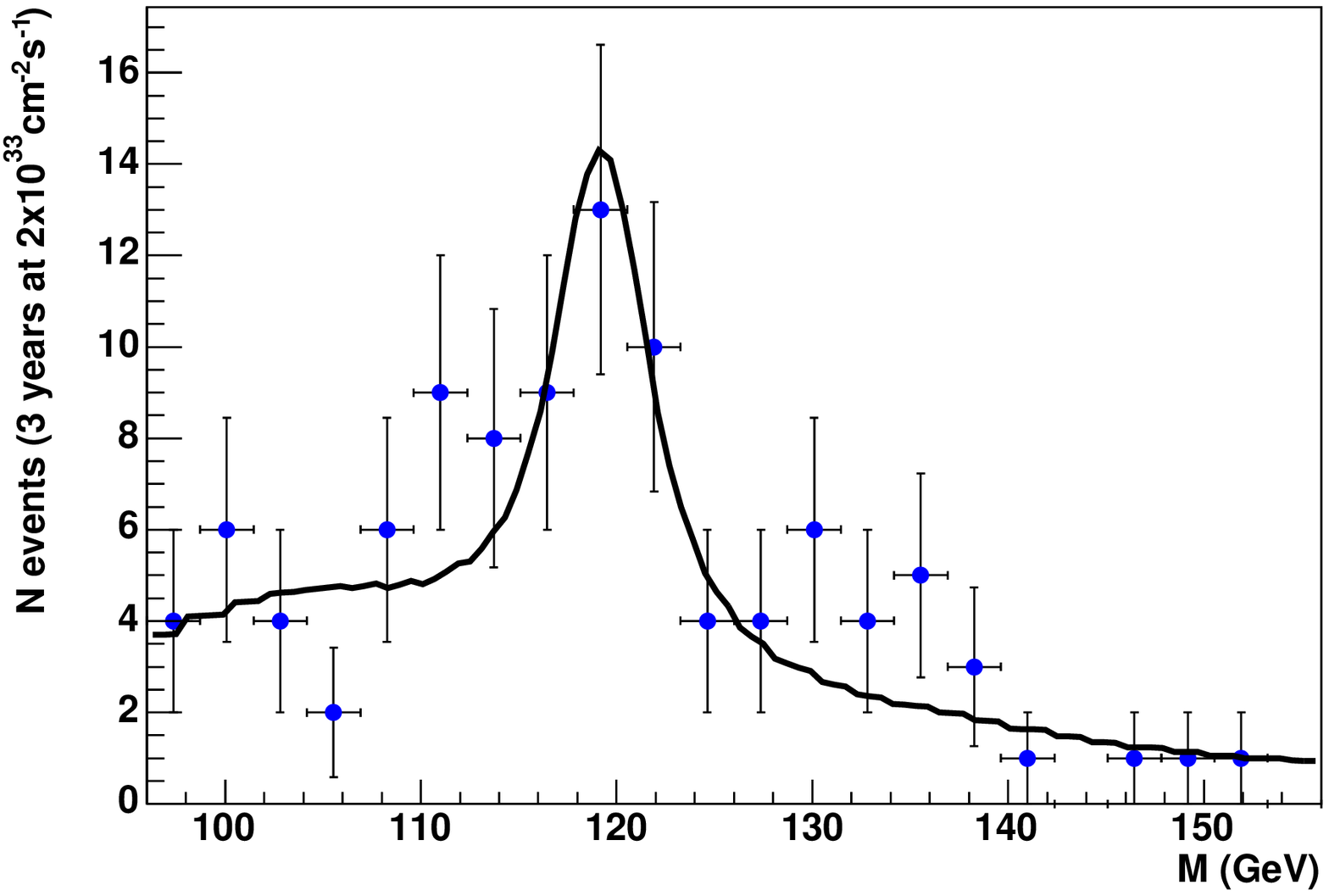,height=1.8in} &
\epsfig{figure=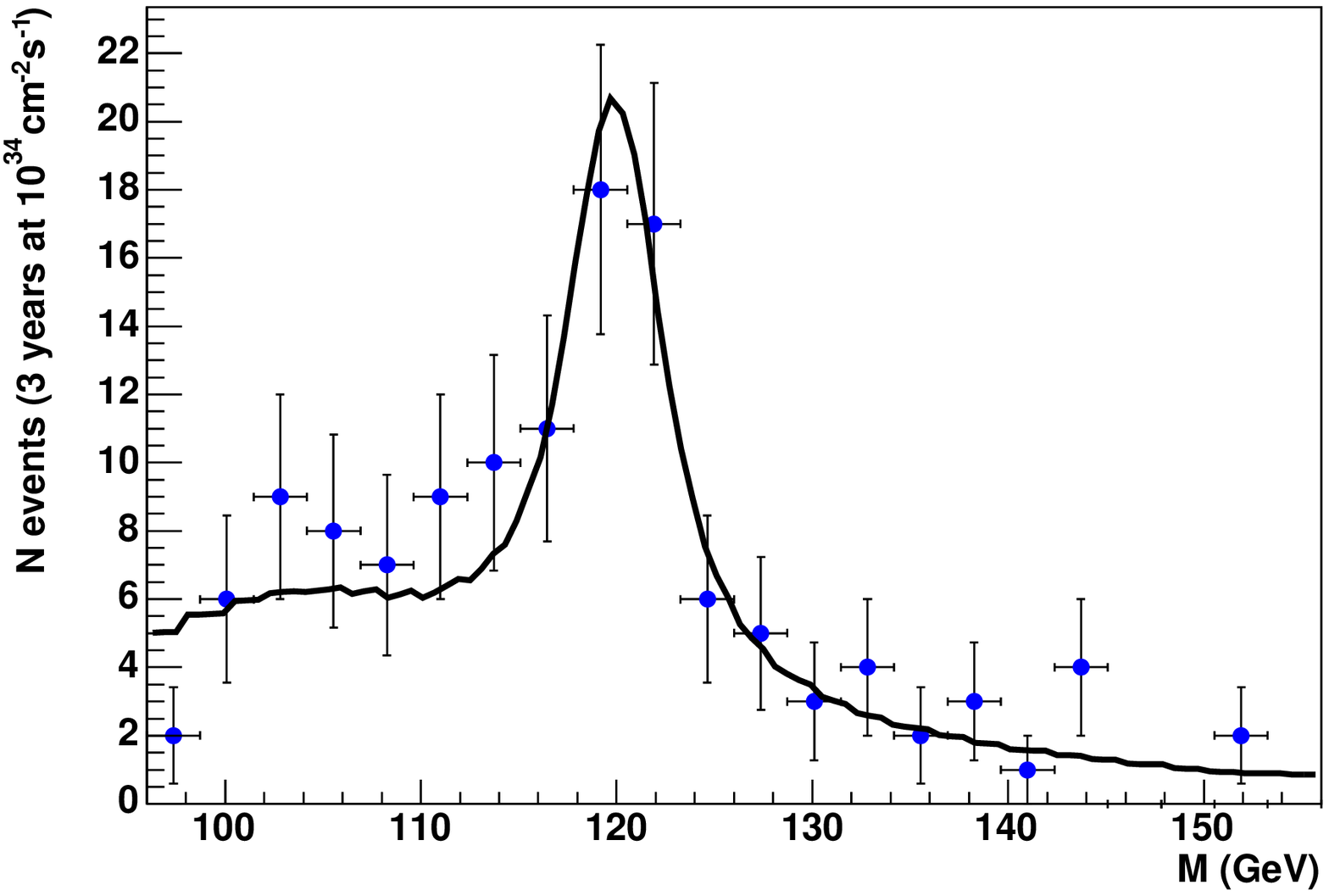,height=1.8in} \\
\end{tabular}
\caption{Higgs signal and background obtained for MSSM Higgs production. 
The signal significance is larger than 3.5 $\sigma$ for 60 fb$^{-1}$
(left plot) and larger than 5 $\sigma$ after three years of data taking at high
luminosity at the LHC and using timing detectors with a resolution of 2 ps
(right plot).}
\label{signif}
\end{center}
\end{figure}

More extensive studies including pile up effects and all background sources 
were performed
recently~\cite{higgsnew} to study in more detail the signal over background for
MSSM Higgs production. The ratio $R$ of the number of diffractive Higgs bosons in
MSSM to SM are given in Fig.~\ref{marek}. We notice that almost the full plane
in ($\tan \beta$, $M_A$) can be covered (typically if $R>$10, the number of
events should be enough to be detected using the diffractive production).
In Fig.~\ref{signif}, we give the number of  background and MSSM Higgs signal events
for a Higgs mass of 120 GeV for $\tan \beta \sim$40.
The signal significance is larger than 3.5 $\sigma$ for 60 fb$^{-1}$
(see Fig.~\ref{signif} left) and larger than 5 $\sigma$ after three years of data taking at high
luminosity at the LHC and using timing detectors with a resolution of 2 ps
(see Fig.~\ref{signif} right).

\subsection{Exclusive top, stop and $W$ pair production at the LHC}
In the same way that Higgs bosons can be produced exclusively, it is possible 
to produce $W$, top and stop quark pairs.
$WW$ bosons are produced via QED processes which means that their cross section
is well determined. On the contrary, top and stop pair production are obtained via
double pomeron exchanges and the production cross section is still uncertain.

The method to reconstruct the mass of heavy objects
double diffractively produced at the LHC is
based on a fit to the turn-on point of the missing mass distribution at 
threshold~\cite{jochen}.

The threshold scan is directly sensitive to
the mass of the diffractively produced object (in the $WW$ case for instance, it
is sensitive to twice the $WW$ mass). The idea is thus to fit the turn-on
point of the missing mass distribution which leads directly to the mass 
of the produced object, the $WW$ boson.

The precision of the $WW$ mass measurement (0.3 GeV for 300 fb$^{-1}$) is not 
competitive with other 
methods, but provides a precise check of the calibration 
of the roman pot detectors. $WW$ events will also allow to assess directly the
sensitivity to the photon anomalous coupling since it would reveal itself by a
modification of the well-known QED $WW$ production cross section. We can notice
that the $WW$ production cross section is proportional to the fourth power of
the $\gamma W$ coupling which ensures a very good sensitivity of that process
\cite{olda}.
The precision of
the top mass measurement might however be competitive, with an expected precision 
better than 1 GeV at high luminosity provided that the cross section
is high enough. 
The other application is to use the so-called ``threshold-scan method"
to measure the stop mass \cite{lavignac}. After taking into account the stop width, we obtain a resolution
on the stop mass of 0.4, 0.7 and 4.3 GeV for a stop mass of 174.3, 210 and 393
GeV for a luminosity (divided by the signal efficiency) of 100 fb$^{-1}$
provided that the cross section is high enough. 

The caveat is  of course that the production via diffractive 
exclusive processes is model dependent --- the production cross section
dependence on mass of the produced object depends on the models --- 
and definitely needs
the Tevatron and LHC data to test the models. It will allow us to determine more precisely 
the production cross section by testing and measuring at the Tevatron the jet 
and photon production for high masses and high dijet or diphoton mass fraction.

\section{The AFP and FP420 projects at the LHC}

\subsection{Motivation}

The motivation to install forward detectors at in ATLAS and CMS is quite
clear. It extends nicely the project of measuring the total cross sections
in ATLAS and TOTEM by measuring hard diffraction at high
luminosity at the LHC. Two locations for the forward detectors are considered at
220 and 420 m respectively to ensure a good coverage in $\xi$ or in mass of the
diffractively produced object as we will see in the following. Installing
forward detectors at 420 m is quite challenging since the detectors will be
located in the cold region of the LHC and the cryostat has to be modified to
accomodate the detectors. In addition, the space available is quite small and
some special mechanism called movable beam pipe are used to move the detectors
close to the beam when the beam is stable enough. The situation at 220 m is
easier since it is located in the warm region of the LHC and both roman pot
and movable beam pipe technics can be used. The AFP (ATLAS Forward Physics)
project is under discussion in the ATLAS collaboration and includes both 220
and 420 m detectors on both sides of the main ATLAS detector.

The physics motivation of this project corresponds to different domains of
diffraction which we already discussed in these lectures:
\begin{itemize}
\item A better understanding of the inclusive diffraction mechanism at the LHC by studying
in detail the structure of pomeron in terms of quarks and gluons as it was done
at HERA~\cite{f2d,us}. Of great importance is also the measurement of the exclusive production
of diffractive events~\cite{oldab} and its cross
section in the jet channel as a function of jet transverse momentum. Its
understanding is necessary to control the background to Higgs signal.
\item Looking for Higgs boson diffractive production in double pomeron exchange in
the Standard Model or supersymmetric extensions of the Standard Model~\cite{ushiggs,
lavignac}. This is
clearly a challenging topic especially at low Higgs boson masses where the
Higgs boson decays in $b \bar{b}$ and the standard non-diffractive search is
difficult. We will detail in the following the trigger strategy. 
\item Sensitivity to the anomalous coupling of the photon by measuring the QED
production cross section of $W$ boson pairs~\cite{olda}. This is one of the best
ways to
access the anomalous coupling before the start of the ILC. Photoproduction of
jets can also be studied.
\%\vspace{-0.3cm}
\item Other topics such as looking for stop events or measuring the top mass
using the threshold scan method~\cite{jochen} which will depend strongly on
the production cross section.
\end{itemize}

\subsection{Forward detector design and location}
\begin{figure}
\begin{center}
\epsfig{file=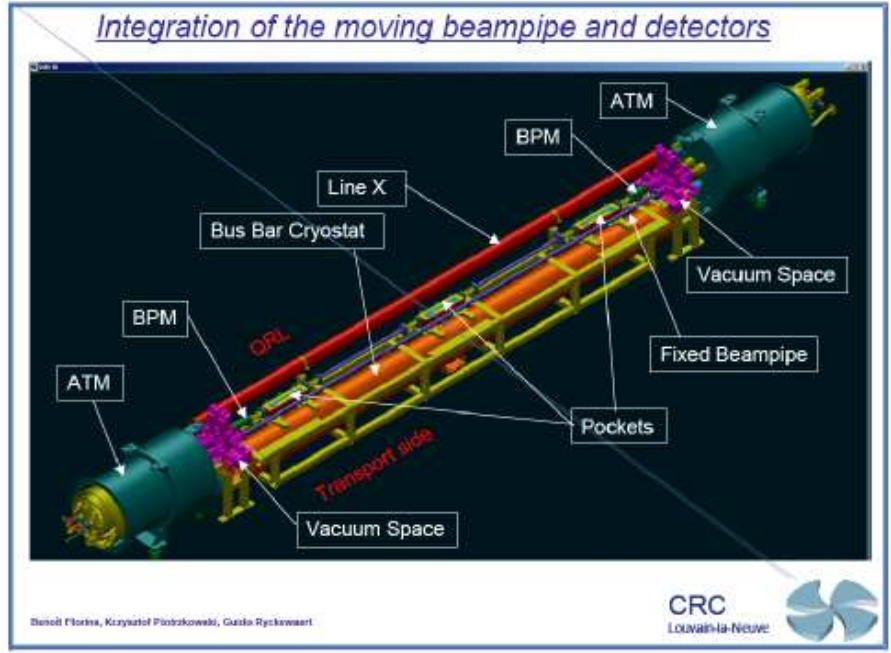,width=12cm}
\caption{Scheme of the movable beam pipe.}
\label{movable}
\end{center}
\end{figure}

As we mentionned in the previous section, it is needed to install movable
beam pipe detectors~\cite{fp420,afp} at 420 m. The scheme of the movable beam pipe
is given in Fig.~\ref{movable}. The principle developed originally for
the ZEUS detector to tag electrons at low angle is quite simple and follows from
the 
same ideas as the roman pots. The beam pipe is larger than the usual one and 
can host the sensitive detectors to tag the diffracted protons in the
final state. When the beam is stable, the beam pipe can move so that the
detectors can be closer to the beam. The movable beam pipe acts in a way as a
single direction roman pot. In Fig.~\ref{movable}, we see the Beam Position
Monitors (BPM) as well as the pockets where the detectors can be put.
The detectors can be aligned and calibrated using the BPMs as well as exclusive
dimuon events.The dimuon mass can be well measured using the central muon
detectors from ATLAS and can be compared to the result obtained using the
missing mass method by tagging the final state proton in the forward detectors.
This allows to calibrate the forward detectors by using data directly.
The exclusive muon production cross section is expected to be high enough to
allow this calibration on a store-by-store basis.

The AFP project proposes to install in addition some forward detectors in ATLAS 
at about 220 m on each side of the main ATLAS detector~\cite{afp}. A few options are still
under discussion and we will discuss only the favoured option at present.
The first component is the
same as at 420 m, namely the movable beam pipe containing the pockets where both
the horizontal Si and timing detectors can be located. However, it is not
possible to use exclusive dimuon events for calibration purposes at 220 m since the cross
section is too small (the acceptance of 220 m detectors is better for high mass
objects, and so higher dimuon masses, which leads to a smaller production
cross section). For calibration and alignment, the idea is to use BPMs as
before and also elastic events. The acceptance in elastic events in the horizontal
detectors in the movable beam pipe is however very small, and this is why 
the idea is to use additional detectors (vertical roman pots) for calibration
purposes. The vertical roman pots can be aligned using the $\xi$ distribution
pointing at 0 for elastics, and the horizontal detectors in the movable beam pipe 
can be calibrated with respect to
the vertical detectors using common events in both detectors (halo or single
diffractive events). With this method, some preliminary studies show that a precision
up to 5 $\mu$m on calibration using elastics is within reach.
The roman pot design follows as close as possible the design which is currently
used by the TOTEM collaboration and the Luminosity group of the ATLAS
collaboration. Another design would be to have only the timing detectors in the
movable beam pipes and three-arm roman pots holding the horizontal and vertical
3D Si detectors.

\begin{figure}
\begin{center}
\epsfig{file=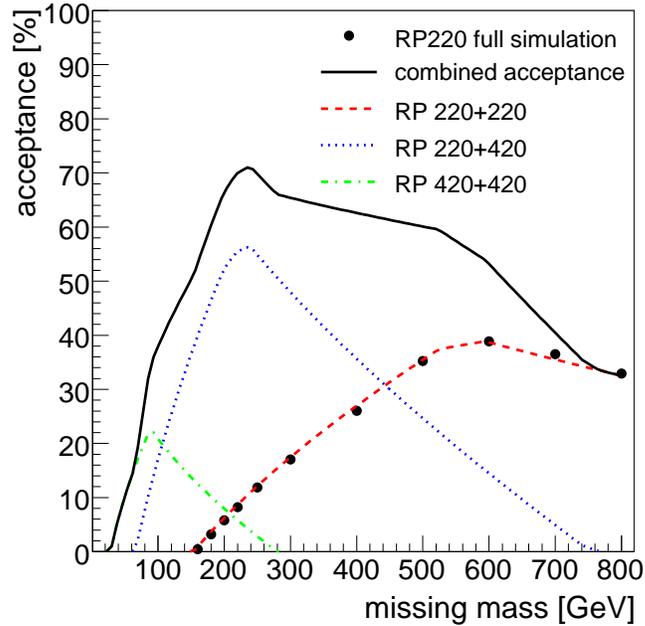,width=9cm}
\caption{Roman pot detector acceptance as a function of missing mass
assuming a 10$\sigma$ operating positions, a dead edge for the detector of 50
$\mu m$ and a thin window of 200 $\mu m$.}
\label{accept1}
\end{center}
\end{figure}

The missing mass acceptance is given in Fig.~\ref{accept1}. 
The missing mass acceptance using only
the 220 m pots starts at 135 GeV, but increases slowly as a function of missing
mass. It is clear that one needs both detectors at 220 and 420 m to obtain a good
acceptance on a wide range of masses since most events are asymmetric (one tag
at 220 m and another one at 420 m). The precision on mass reconstruction using
either two tags at 220 m or one tag at 220 m and another one at 420 m is of the
order of 2-4 \% on the full mass range, whereas it goes
down to 1\% for symmetric 420 m tags.

\subsection{Detectors inside forward detectors for the AFP project}
We propose to put inside the forward detectors two kinds of detectors, 
namely 3D Silicon
detectors to measure precisely the position of the diffracted protons, and the mass
of the produced object and
$\xi$, and precise timing detectors.

The position detectors will consist in 3D Silicon
detector which allow to obtain a resolution in position better than
10 $\mu$m. The detector is made of 10 layers of 3D Si pixels of 50 $\times$
400 $\mu$m. One layer contains 9 pairs of columns of 160 pixels,
the total size being 7.2 $\times$ 8 mm$^2$. 
The detectors will
be read out by the standard ATLAS pixel chip~\cite{fp420,afp}. The latency time of the chip is 
larger than 6 $\mu$s which gives enough time to send back the local L1 decision from
the roman pots to ATLAS (see the next paragraph about trigger for more detail),
and to receive the L1 decision from ATLAS, which means a distance of about 440
m. It is also foressen to perform a slight modification of the chip to
include the trigger possibilities into the chip.
It is planed to install the roman pot together with the Silicon detectors during
a shut down of the LHC in 2010.

The timing detectors are necessary at the highest luminosity of the LHC to
identify from which vertex the protons are coming from. It is expected that up
to 35 interactions occur at the same bunch crossing and we need to identify from
which interaction, or from which vertex the protons are coming from. A precision
of the order of 1 mm or 2-5 ps is required to distinguish between the
different vertices and to make sure that the diffracted protons come from the
hard interactions. Picosecond timing detectors are still a challenge and are
developped for medical and particle physics applications. 
Two technogies are developped, either using as a radiator --- with the aim
to emit photons by the diffracted protons ---- gas(gas Cerenkov detector or GASTOF) 
or a crystal of about 2.5 cm (QUARTIC), and the signal can be read out by 
Micro-Channel Plates Photomultipliers
developped by Photonis~\cite{fp420,afp}. The space resolution of those detectors should be of the
order of a few mm since at most two protons will be detected in those detectors
for one given bunch crossing at the highest luminosity. The detectors can be read
out with a Constant Fraction Discriminator which allows to improve the timing
resolution significantly compared to usual electronics.
A first version of the
timing detectors is expected to be ready in 2010 with a resolution of
20-30 ps, and the final version by 2012-2013 with a resolution of 2-5 ps.

\subsection{Trigger principle and rate}
In this section, we would like to give the principle of the trigger using the
roman pots at 220 m as well as the rates obtained using a simulation of the
ATLAS detector and trigger framework~\cite{afp}.

\begin{figure}
\begin{center}
\epsfig{file=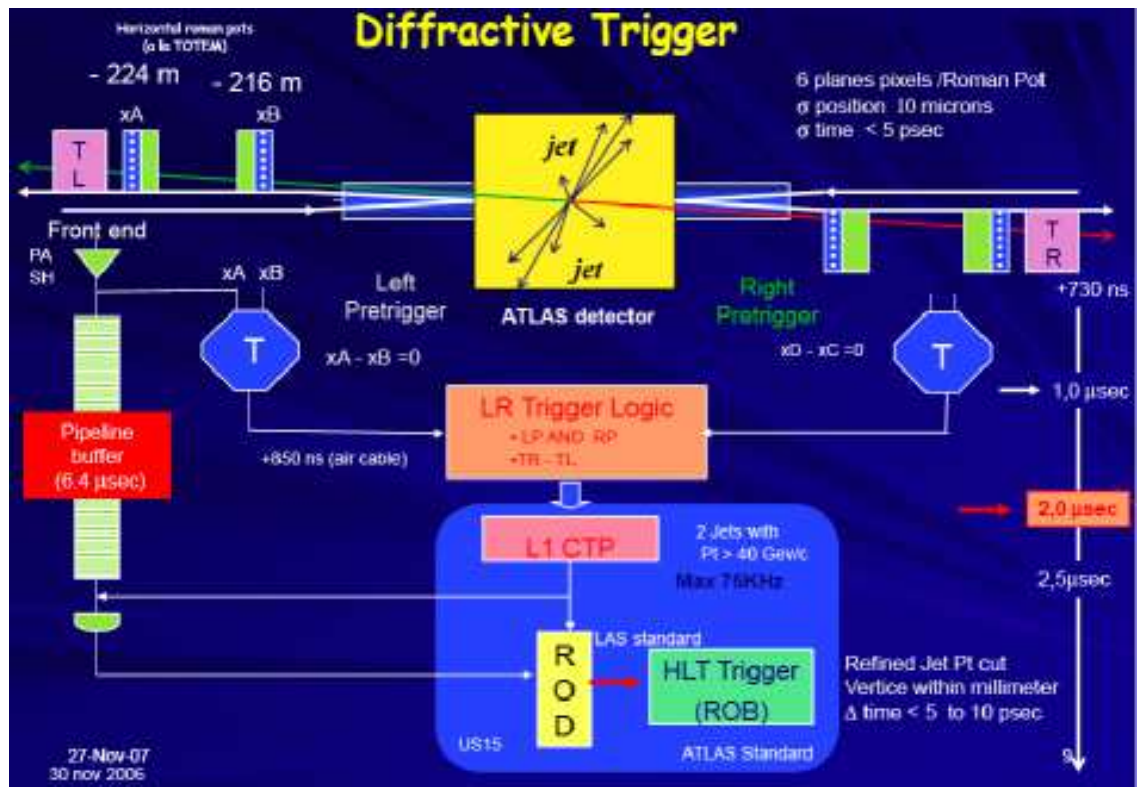,width=15cm}
\caption{Scheme for L1 trigger for the AFP project.}
\label{trig}
\end{center}
\end{figure}

The principle of the trigger is shown in Fig.~\ref{trig} in the case of a Higgs boson
decaying into $b \bar{b}$ as an example.
The first level trigger comes directly from two different 3D Silicon layers
in each forward detector. It is more practical to use two dedicated planes for
triggering only since it allows to use different signal thresholds for trigger
and readout. The idea is to send at most five strip addresses which are hit
at level 1 (to simplify the trigger procedure, we group all pixels in 
vertical lines as one element only for the trigger
since it is enough to know the distance in the horizontal direction to have 
a good approximation of $\xi$). 
A local trigger is defined at the roman pot level on each side of
the ATLAS experiment by combining the two trigger planes in each roman pot and
the roman pots as well. If the hits are found to be compatible (not issued by
noise but by real protons), the strip addresses are sent to ATLAS, which allows
to compute the $\xi$ of each proton, and the diffractive mass. This information
is then combined with the information coming from the central ATLAS detector,
requesting for instance two jets above 40 GeV in the case shown in Fig.~\ref{trig}. At
L2, the information coming from the timing detectors for each diffracted proton
can be used and combined with the position of the main vertex of ATLAS to check
for compatibility. Once a positive ATLAS trigger decision is taken (even without
any diffracted proton), the readout informations coming from the 
roman pot detectors are sent to ATLAS as any subdetector.

The different trigger possibilities for the roman pots are given below:
\begin{itemize}
\item {\it {\bf Trigger on DPE events at 220 m:}}
This is the easiest situation since two protons can be requested at Level 1 
at 220 m. Three different options are considered:
\newline
{\it - trigger on high mass Higgs} ($M>160$ GeV) given by ATLAS directly
(decay in $WW$, $ZZ$),
\newline
{\it - inclusive trigger on high mass object} by requesting two high $p_T$ jets
and two positive tags in roman pots,
\newline 
{\it - trigger on jets} (high $p_T$ jets given directly by ATLAS, and low $p_T$
jet special trigger for QCD studies highly prescaled).
\newline
This configuration will not rise any problem concerning the L1 rate since most
of the events will be triggered by ATLAS anyway, and the special diffractive
triggers will be for QCD measurements and can be highly prescaled.

\item {\it {\bf Trigger on DPE events at 220 and 420 m:}}
This is the most delicate scenario since the information from the 420 m pots cannot
be included at L1 because of the L1 latency time of ATLAS. 
The strategy is the following (see Table 1):
\newline
{\it - trigger on heavy objects} (Higgs...) decaying in $b \bar{b}$ by requesting
a positive tag (one side only) at 220 m with $\xi < 0.05$
(due to the $420\mathrm{m}$ RP acceptance
in  $\xi$, the proton momentum fractional loss in the $220\mathrm{m}$ roman pot
cannot be too high
if the Higgs mass is smaller than $140\,\mathrm{GeV}$), and topological cuts
on jets such as the exclusiveness of the process
($(E_{jet1}+E_{jet2})/E_{calo}>0.9$, 
$(\eta_1+\eta_2)\cdot\eta_{220} > 0$, where 
$\eta_{1,2}$ are the pseudorapidities of the two L1 jets, and $\eta_{220}$
the pseudorapidity of the proton in the $220\mathrm{m}$ roman pots). 
This trigger can hold without prescales to a
luminosity up to 2.10$^{33}$ cm$^{-2}$s$^{-1}$,
\newline
{\it - trigger on jets} (single diffraction, or double pomeron exchange) for QCD
studies: can be heavily prescaled,
\newline
{\it - trigger on $W$, top...} given by ATLAS with lepton triggers.
\newline
Let us note that the rate will be of the order of a few Hz at L2 by adding a cut on
a presence of a tag in the 420 pots, on timing, and also on the compatibility of
the rapidity of the central object computed using the jets or the protons in
roman pots.
\end{itemize}

\begin{footnotesize}
\begin{table}
\begin{center}
\begin{tabular}{|c|c|c|c|c|c|}
\hline
${\cal L}$ & $n_{pp}$ per & 2-jet  & RP200 
& $\xi < 0.05$  & Jet \cr
$E_T > 40\,\mathrm{GeV}$ & bunch & rate $[\mathrm{kHz}]$ &
reduction & reduction & Prop.  \cr
 & crossing & $[\mathrm{cm}^{-2}\cdot\mathrm{s}^{-1}]$& factor & factor & \cr
\hline
$1\times10^{32}$ & 0.35 & 2.6 & 120 & 300 & 1200 \cr
$1\times10^{33}$ & 3.5 & 26 & 8.9 & 22 & 88 \cr
$2\times10^{33}$ & 7 & 52 & 4.2 & 9.8 & 39.2 \cr
$5\times10^{33}$ & 17.5 & 130 & 1.9 & 3.9 & 15.6 \cr
$1\times10^{34}$ & 35 & 260 & 1.3 & 2.2 & 8.8 \cr
\hline
\end{tabular}
\end{center}
\caption{L1 rates for 2-jet trigger with $E_T > 40\,\mathrm{GeV}$ and
additional reduction factors due to the requirement of triggering on
diffractive proton at $220\,\mathrm{m}$, and also on jet properties.}
\label{t_trigger}
\end{table}
\end{footnotesize}

\section{Conclusion}
In these lectures, we presented and discussed the most recent results on
inclusive diffraction from the Tevatron experiments and gave the
prospects for the future at the LHC. Of special interest is the exclusive
production of Higgs boson and heavy objects ($W$, top, stop pairs) which will
require a better understanding of diffractive events and the link between $ep$
and hadronic colliders, and precise measurements and analyses of inclusive
diffraction at the LHC in particular to constrain further the gluon density in
the pomeron. The search for exclusive events at the Tevatron is quite promising
especially in the dijet channel. 
We finished the lectures by describing the main projects to install
forward detectors at the LHC, and especially the AFP project to measure diffraction
at high luminosity.

\section*{Acknowledgments}
I thank Robi Peschanski for a careful reading of the manuscript.



\begin{thebibliography}{10}

\bibitem{loehr} B. Loehr, contribution about ``Diffraction at HERA" at this
Summer school

\bibitem{chic} 
M. Rangel, C. Royon, G. Alves, J. Barreto, R. Peschanski, Nucl. Phys. B{\bf 774}
(2007) 53; M. Rangel these proceedings.


\bibitem{oldtev} CDF Collaboration, Phys. Rev. Lett. {\bf 87} (2001) 241802;
Phys. Rev. Lett. {\bf 78} (1997) 2698; D0 Collaboration
Phys. Lett.  {\bf B574} (2003) 169; Phys. Lett. {\bf B 531} (2002) 52.


\bibitem{cdfpots} 
See http://www-cdf.fnal.gov and http://rock23.rockefeller.edu/.

\bibitem{d0pots} 
Proposal for a Forward 
Proton Detector at D\O\ , D\O\
Collaboration (1997), Proposal P-900 to FERMILAB PAC.

\bibitem{f2d} H1 Collaboration, Eur. Phys.J. C{\bf 48} (2006) 715;
H1 Collaboration,  Eur. Phys.J. C{\bf 48} (2006) 749;
H1 Collaboration, Z. Phys. C{\bf 76} (1997) 613; ZEUS
Collaboration,
Nucl. Phys. B{\bf 713} (2005) 3.



\bibitem{bekw} 
J.Bartels, J.Ellis, H.Kowalski, M.Wuesthoff,  
Eur.Phys.J.C7 (1999) 443,
J.Bartels, C.Royon, Mod. Phys. Lett. {\bf A14} (1999) 1583. 

\bibitem{dipole} 
A.H.Mueller and B.Patel, 
Nucl. Phys. {\bf B425} (1994) 471; 
A.H.Mueller, Nucl. Phys. {\bf B437} (1995) 107; 
A.H.Mueller, Nucl. Phys. {\bf B415} (1994) 373;
H.Navelet, R.Peschanski, Ch.Royon, S.Wallon,
Phys. Lett. {\bf B385} (1996) 357; A. Bialas, R.Peschanski, C.Royon,
Phys. Rev. {\bf D57} (1998) 6899; 
S.Munier, R.Peschanski, C.Royon, Nucl. Phys. 
{\bf B534} (1998) 297; M. Boonekamp, A. De Roeck, C. Royon, S. Wallon,
Nucl. Phys. {\bf B555} (1999) 540.




\bibitem{saturation} 
K. Golec-Biernat and M. Wusthoff,
Phys.\ Rev. {\bf D59} (1999) 014017;
Phys.\ Rev. {\bf D60} (1999) 114023.


\bibitem{dglap} 
G.Altarelli and G.Parisi,
{\it Nucl. Phys.} {\bf B126}  18C (1977) 298;
V.N.Gribov and L.N.Lipatov, {\it Sov. Journ. Nucl. Phys.} (1972) 438 and 675;
Yu.L.Dokshitzer, {\it Sov. Phys. JETP.} {\bf 46} (1977) 641.

\bibitem{ingelman} G. Ingelman, P.E. Schlein, Phys. Lett. {\bf B152} (1985) 256.

\bibitem{us} 
C. Royon, L. Schoeffel, J. Bartels, H. Jung, R. Peschanski, Phys. Rev. {\bf D63}
(2001) 074004;
C. Royon, L. Schoeffel, R. Peschanski, E. Sauvan, 
Nucl. Phys. B{\bf 746} (2006) 15;
C. Royon, L. Schoeffel, S. Sapeta, R. Peschanski, E. Sauvan, 
Nucl. Phys. B{\bf 781} (2007) 1. 


\bibitem{collins} 
J. Collins, Phys. Rev. D{\bf 57} (1998) 3051.



\bibitem{cdfdiff} 
M. Gallinaro, talk given at the DIS 2006
workshop, 20-24 April 2006, Tsukuba, Japan, see /http://www-conf.kek.jp/dis06/; 
Dino Goulianos, talk given at the low x 2006 workshop, June 28 - July 1 2206,
Lisbon, Portugal, 
see http://www-d0.fnal.gov/~royon/lowx\_lisbon; CDF Collaboration,
Phys. Rev. Lett. {\bf 88} (2002) 151802.

\bibitem{cdffact} CDF Collaboration, Phys. Rev. Lett. {\bf 84}
(2000) 5043; Phys. Rev. Lett. {\bf 87} (2001) 141802.

\bibitem{alexander} A. Kup\v{c}o, R. Peschanski, C.Royon,
Phys. Lett. {\bf B606} (2005) 139, and references therein. 


\bibitem{sci} 
A. Edin, G. Ingelman, J. Rathsman, Phys. Lett. B{\bf 366} (1996) 371. 


\bibitem{cdfchic} 
CDF Collaboration, analysis and results described in: http://www-cdf.fnal.gov/physics/new/qcd/abstracts/dpe\_ex\_07.html.


  
\bibitem{cdfgamma} CDF Collaboration, Phys. Rev. Lett. 99 (2007) 242002. 

\bibitem{cdfrjj} CDF Collaboration, preprint hep-ex/0712.0604.


\bibitem{oldab} O.Kepka, C. Royon, Phys. Rev.D{\bf 76} (2007) 034012; O. Kepka,
C. Royon these proceedings.

\bibitem{ushiggs}   C.~Royon,
  Mod.\ Phys.\ Lett.\ A {\bf 18}, 2169 (2003) and references therein;
M. Boonekamp, R. Peschanski, C. Royon, Phys. Rev. Lett. {\bf  87 } 
(2001) 
251806; Nucl. Phys. {\bf B669} (2003) 277;
M. Boonekamp, A. De Roeck, R. Peschanski, C. Royon, Phys. Lett.  {\bf  
B550} (2002) 93;
V.A. Khoze, A.D. Martin, M.G. Ryskin, Eur. Phys. J. {\bf C19} (2001) 477;
Eur. Phys. J. {\bf C23} (2002) 311;
Eur. Phys. J. {\bf C24} (2002) 581; arXiv:0802.0177;
Phys. Lett. B650 (2007) 41; A.B. Kaidalov, V.A. Khoze, A.D. Martin, M.G. Ryskin,
Eur. Phys. J. {\bf C33} (2004) 261; Eur. Phys. J. {\bf C31} (2003) 387


\bibitem{tev4lhc} C. Royon, preprint Fermilab-CONF-06-018E,
hep-ph/0601226; TeV4LHC QCD Working Group, FERMILAB-CONF-06-359, hep-ph/0610012.


\bibitem{atlaslumi} 
ATLAS Coll., see
http://atlas-project-lumi-fphys.web.cern.ch/atlas-project-lumi-fphys/;
C. Royon, Proccedings of Science DIFF2006 (2006) 021. 


\bibitem{totem} 
TOTEM Coll., see http://totem.web.cern.ch/Totem/, TOTEM Technical Design Report,

\bibitem{tevtotal} E710 Collaboration, Phys. Rev. Lett. {\bf 63} (1989) 2784;
E811 Collaboration, Phys. Lett. B{\bf 445} (1999) 419; CDF Collaboration,
Phys. Rev. D{\bf 50} (1994) 5550.


\bibitem{ua4} UA4 Coll., Phys. Lett. B{\bf 198} (1987) 583.



\bibitem{fp420} 
FP420 Coll., see http://www.fp420.com

\bibitem{afp} AFP TDR in ATLAS to be submitted; see: 
http://project-rp220. web.cern.ch/project-rp220/index.html; C. Royon,
preprint arXiv:0706.1796, proceedings of 15th International 
Workshop on Deep-Inelastic Scattering and Related Subjects (DIS2007), 
Munich, Germany, 16-20 Apr 2007. 

\bibitem{dpemc} See http://boonekam.home.cern.ch/boonekam/dpemc.htm


\bibitem{lavignac} 
M. Boonekamp, J. Cammin, S. Lavignac, R. Peschanski, C. Royon,
Phys. Rev. {\bf D73} (2006) 115011, and references therein.


\bibitem{higgsnew} B. Cox, F. Loebinger, A. Pilkington, JHEP 0710 (2007) 090;
S. Heinemeyer et al., Eur.~Phys.~ J. C {\bf 53} (2008) 231.

\bibitem{jochen} 
M. Boonekamp, J. Cammin, R. Peschanski, C. Royon, 
Phys. Lett. B{\bf 654} (2007) 104.


\bibitem{olda} 
O. Kepka, C. Royon in preparation.



\end{thebibliography}
\end{document}